\documentclass[aps, twocolumn,superscriptaddress, 
amsmath,amssymb,reprint,numbers,noeprint, floatfix,longbibliography]{revtex4-1}

\usepackage[english]{babel}
\usepackage{graphicx}
\usepackage{amsmath}
\usepackage{physics}
\usepackage{amssymb}
\usepackage{physics}

\begin{document}

\title{Hybrid Integration of GaP Photonic Crystal Cavities with Silicon-Vacancy Centers in Diamond by Stamp-Transfer}

\author{Srivatsa Chakravarthi}
\affiliation{University of Washington, Physics Department, Seattle, WA, 98105, USA}%
\author{Nicholas S. Yama}
\affiliation{University of Washington, Electrical and Computer Engineering Department, Seattle, WA, 98105, USA}%
\author{Alex Abulnaga}
\affiliation{Princeton University, Electrical and Computer Engineering Department, New Jersey 08544, USA}
\author{Ding Huang}
\affiliation{Princeton University, Electrical and Computer Engineering Department, New Jersey 08544, USA}%
\author{Christian Pederson}
\affiliation{University of Washington, Physics Department, Seattle, WA, 98105, USA}%
\author{Karine Hestroffer}
\affiliation{Department of Physics, Humboldt-Universitat zu Berlin, Newtonstrasse, Berlin, Germany}
\author{Fariba Hatami}
\affiliation{Department of Physics, Humboldt-Universitat zu Berlin, Newtonstrasse, Berlin, Germany}
\author{Nathalie P. de Leon} 
\affiliation{Princeton University, Electrical and Computer Engineering Department, New Jersey 08544, USA}
\author{Kai-Mei C. Fu}%
\affiliation{University of Washington, Electrical and Computer Engineering Department, Seattle, WA, 98105, USA}%
\affiliation{University of Washington, Physics Department, Seattle, WA, 98105, USA}%
\affiliation{Physical Sciences Division, Pacific Northwest National Laboratory, Richland, Washington 99352, USA}

\date{\today}

\begin{abstract}
\vspace{2em}

Optically addressable solid-state defects are emerging as one of the most promising qubit platforms for quantum networks. Maximizing photon-defect interaction by nanophotonic cavity coupling is key to network efficiency. We demonstrate fabrication of gallium phosphide 1-D photonic crystal waveguide cavities on a silicon oxide carrier and subsequent integration with implanted silicon-vacancy (SiV) centers in diamond using a stamp-transfer technique. The stamping process avoids diamond etching and allows fine-tuning of the cavities prior to integration. After transfer to diamond, we measure cavity quality factors ($Q$) of up to 8900 and perform resonant excitation of single SiV centers coupled to these cavities. For a cavity with $Q$ of 4100, we observe a three-fold lifetime reduction on-resonance, corresponding to a maximum potential cooperativity of $C = 2$. These results indicate promise for high photon-defect interaction in a platform which avoids fabrication of the quantum defect host crystal. 

\vspace{4em}
\end{abstract}

\maketitle

In an optical quantum network, quantum entanglement is distributed among spatially separated stationary qubit nodes through flying qubits (photons)~\cite{ruf2021quantum}, enabling long-range quantum communication~\cite{wehner2018quantum} and distributed quantum computing~\cite{ref:kielpinski2002als}. Optically addressable defects~\cite{bassett2019quantum} in materials like diamond~\cite{doherty_nitrogen-vacancy_2013,bradac_quantum_2019}, silicon carbide~\cite{radulaski_scalable_2017,miao2020universal}, ZnO~\cite{viitaniemi2022coherent} and rare-earth doped crystals~\cite{zhong_nanophotonic_2017,raha2020optical} show tremendous potential for realization of such stationary qubit nodes. Efficient generation and utilization of flying qubits can be enabled by on-chip photonic integration of defect qubits. Here, we demonstrate a hybrid-photonic architecture in diamond, integrating single negatively charged silicon-vacancy qubits with waveguide integrated gallium phosphide (GaP) photonic crystal (PhC) cavities~\cite{Huang21}.

\begin{figure*}[ht]
\centering
\includegraphics[width=\textwidth]{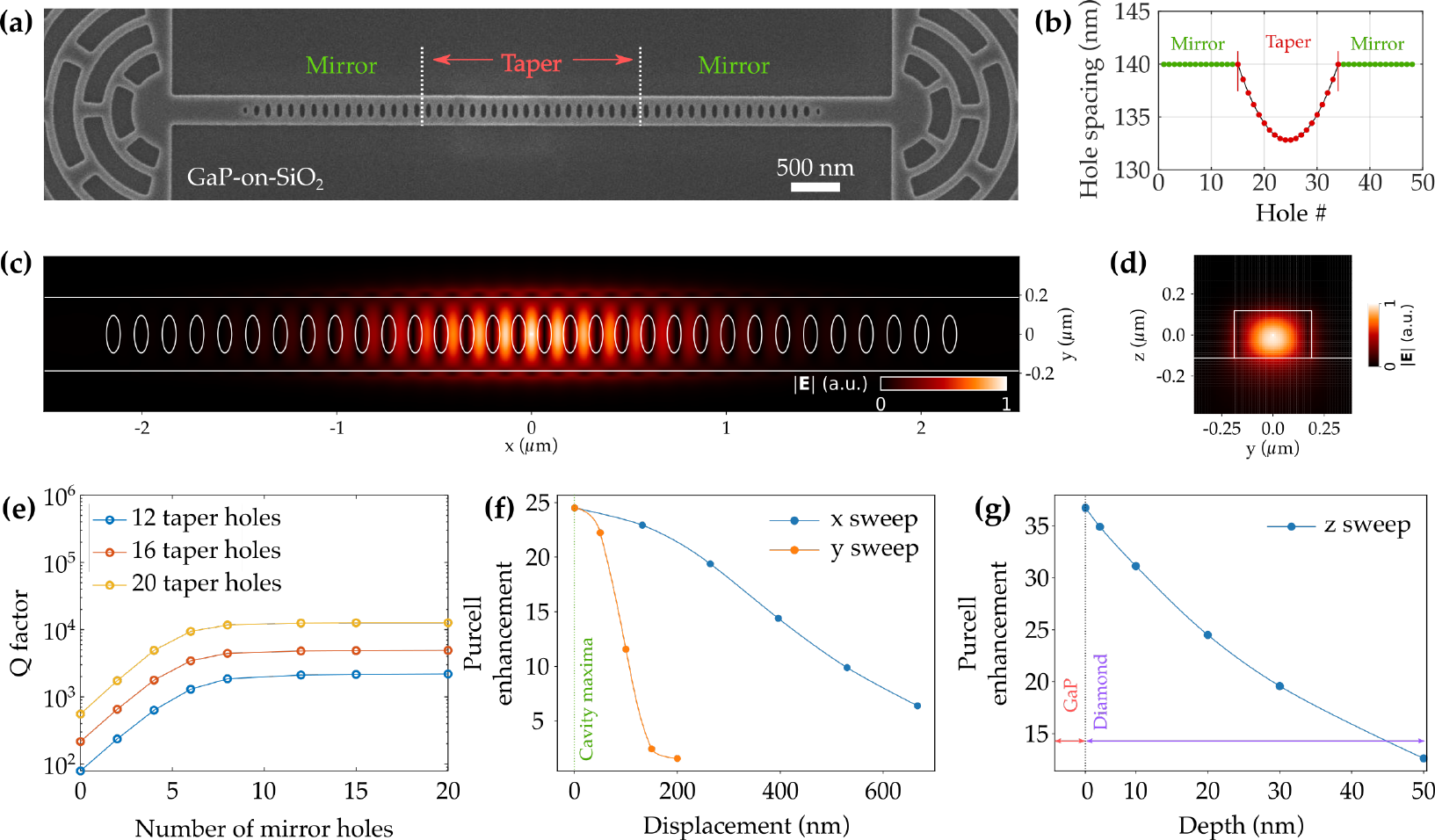}
\caption{(a) SEM image of a GaP 1-D cavity. (b) The variation of the hole spacing across the cavity, distinguishing the mirror and taper regions. (c, d) Simulated E-field intensity of the primary cavity mode across the xy-plane ((c), z$=$0 centered on GaP layer) and yz-plane ((d), x$=$0 centered on cavity maxima) within the GaP-on-diamond structure. (e) Simulated scaling of the total $Q$-factor of the GaP-on-diamond cavity with increasing number of mirror and taper holes. (f, g) Simulated Purcell enhancement for a $<$111$>$ oriented dipole at z$=$20\,nm below the diamond surface as a function of displacement from the cavity maxima along the xy-plane (f) and at $x=y=0$ as a function of the depth of the dipole within the diamond host. (g). For our fabricated devices, we select a Si implantation at of depth of 20\,nm.
}
\label{fig:phc_design}
\end{figure*}

Negatively charged silicon-vacancy (SiV) centers in high-quality diamond crystals show nearly lifetime-limited optical transitions for centers created by ion-implantation and vacuum annealing~\cite{lang2020long}. The favorable spectral behavior of the SiV center can be attributed to its inversion symmetry which protects the optical transitions from electric field noise in the environment~\cite{hepp2014electronic,sipahigil_indistinguishable_2014}. This protection has enabled efficient coupling of SiV centers to monolithic diamond photonic structures~\cite{sipahigil2016integrated}. Here we explore an alternative planar geometry that eliminates diamond etching, potentially minimizing fabrication-induced damage to the defect environment. Photonic cavities are fabricated on an intermediate GaP on silicon-oxide carrier chip and then transferred to a diamond sample with implanted SiV defects via a polymer stamping process~\cite{dibos_atomic_2018, Huang21}. The hybrid stamp transfer process will allow independent optimization of the SiV qubits and the photonic structures. Commercial wafer scale implementations of stamp-transfer pick-and-place hybrid photonic integration processes~\cite{de2020heterogeneous,heck2022inspire,billet2022gallium} suggest good prospects for scalable assembly of defect-cavity interfaces with ancillary devices such as detectors, switches, filters and passive photonic components.

In our experiments, we demonstrate single SiV qubits coupled to GaP photonic cavities. We observe a quality factors on the silicon oxide carrier exceeding $Q_{\mathrm{SiO_2}}$ $30,000$ and upto $Q_{\mathrm{Dia}}=8,900$ after stamp transfer to the diamond substrate. For a cavity with a $Q_{\mathrm{Dia}}=4,100$ we see a three-fold reduction in the coupled SiV lifetime with the cavity on-resonance. The observed lifetime reduction corresponds to a minimum cavity Purcell enhancement of 30 and maximum possible cooperativity of 2 for the coupled SiV center.

\begin{figure*}[ht]
\centering
\includegraphics[width=\textwidth]{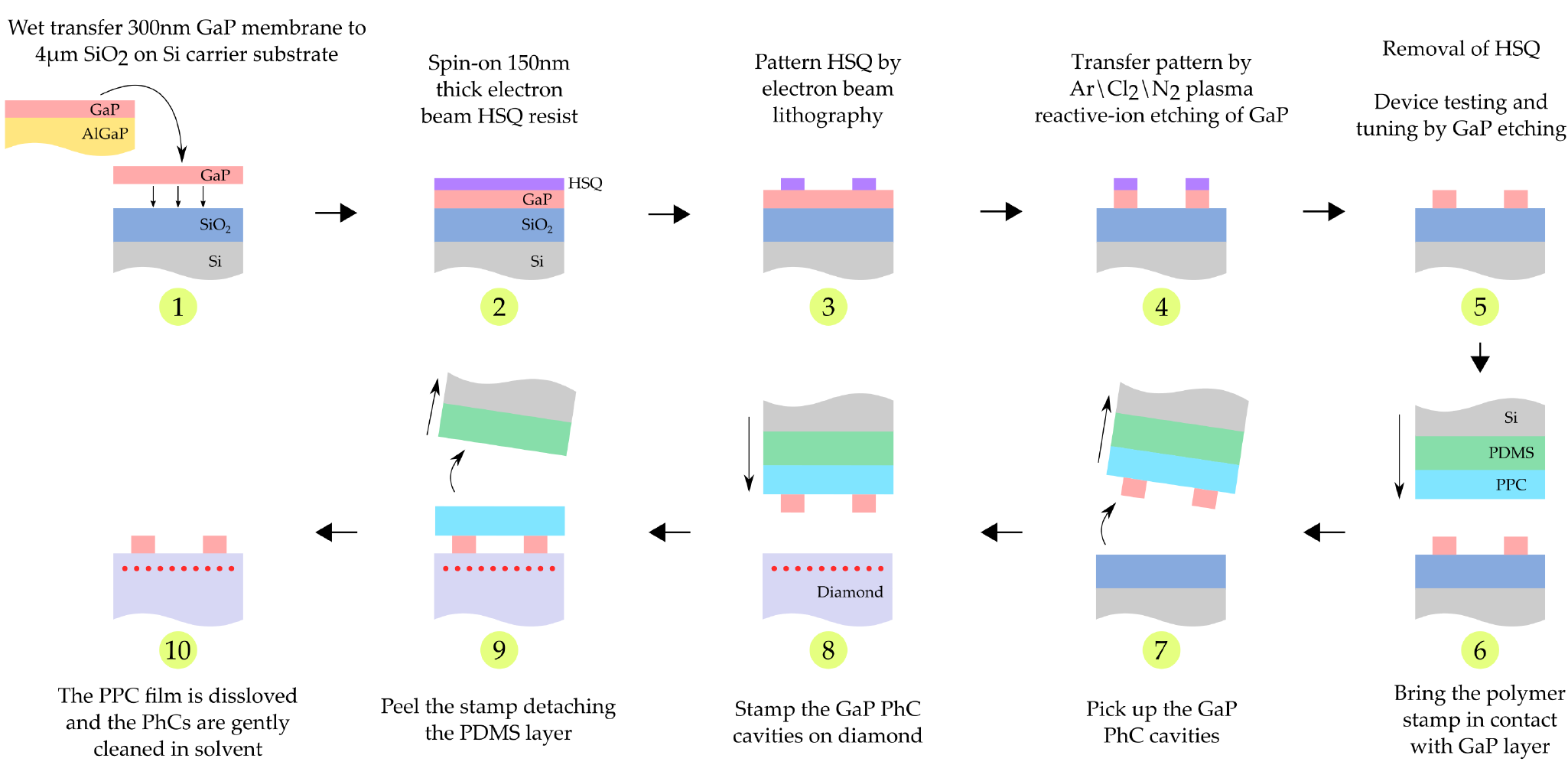}
\caption{An illustration of the stamp transfer process. We begin with GaP cavities fabricated on a silicon oxide carrier chip and finish with cavities integrated with implanted SiV centers in a host diamond chip. Microscope images from various stages of fabrication are provided in SI.2.
}
\label{fig:phc_stamp}
\end{figure*}

The device consists of a 1-D array of elliptical holes etched into the GaP (Fig.~\ref{fig:phc_design}a). The designed GaP layer thickness is constrained to 222\,nm. A unit cell of the cavity consists of a single elliptical hole with major-axis $h_y$ and minor-axis $h_x$, GaP width $w$ and a lattice constant $a_{\mathrm{cav}}$. The of the cavity can be divided into the taper and mirror sections as shown in Fig.~\ref{fig:phc_design}b. Within the mirror region, the hole periodicity, $a_{\mathrm{mir}}$, is chosen to maximize the TE quasi-band gap around the target frequency of the SiV zero-phonon transition at 737\,nm. Within the taper region, the hole periodicity is adiabatically reduced from $a_{\mathrm{mir}}$ so as to create a defect mode for the target frequency. The presence of the high index diamond substrate underneath the GaP introduces design challenges towards achieving high $Q$ because the substrate expands the light-line such that photons can couple to the continuum of modes in the diamond and thus leak out of the cavity. Full details of the cavity design and optimization can be found in Ref.\cite{raha2021telecom,Huang21}. 

We restrict the minimum feature size to be 70\,nm due to fabrication limitations. The optimized design, with $(a_{\mathrm{mir}}, a_{\mathrm{cav}}, h_x, h_y, w) = (140, 133, 71, 183, 360)$\,nm is simulated using finite-difference time-domain method (FDTD, Lumerical). A driven dipole source modeling the SiV is placed within the evanescent cavity field. The time-decay of the cavity mode and measured radiated power is utilized to estimate the resonator characteristics and tolerance to dipole placement (Fig.~\ref{fig:phc_design}c-g, see SI.7 for details). For a defect 20\,nm beneath the GaP-diamond interface a Purcell enhancement as high as 25 is possible, which increases to 35 at 3\,nm beneath the interface.  

The cavity structure is fabricated using a 300\,nm GaP membrane (Fig.~\ref{fig:phc_stamp}:1-5) that is released from a GaP substrate by etching an intermediate $\mathrm{Al_{0.8}Ga_{0.2}P}$ sacrificial layer with dilute HF. The GaP membrane is then transferred to a silicon-oxide on Si substrate using a wet-transfer process~\cite{yablonovitch_van_1990,schmidgall_frequency_2018,chakravarthi_inverse-designed_2020}. Electron-beam lithography is performed using hydrogen silsesquioxane (HSQ) resist to pattern the cavity devices. Subsequent inductively coupled plasma reactive-ion-etching (RIE) of the GaP layer forms the photonic devices.

\begin{figure}[t]
\centering
\includegraphics[width=\columnwidth]{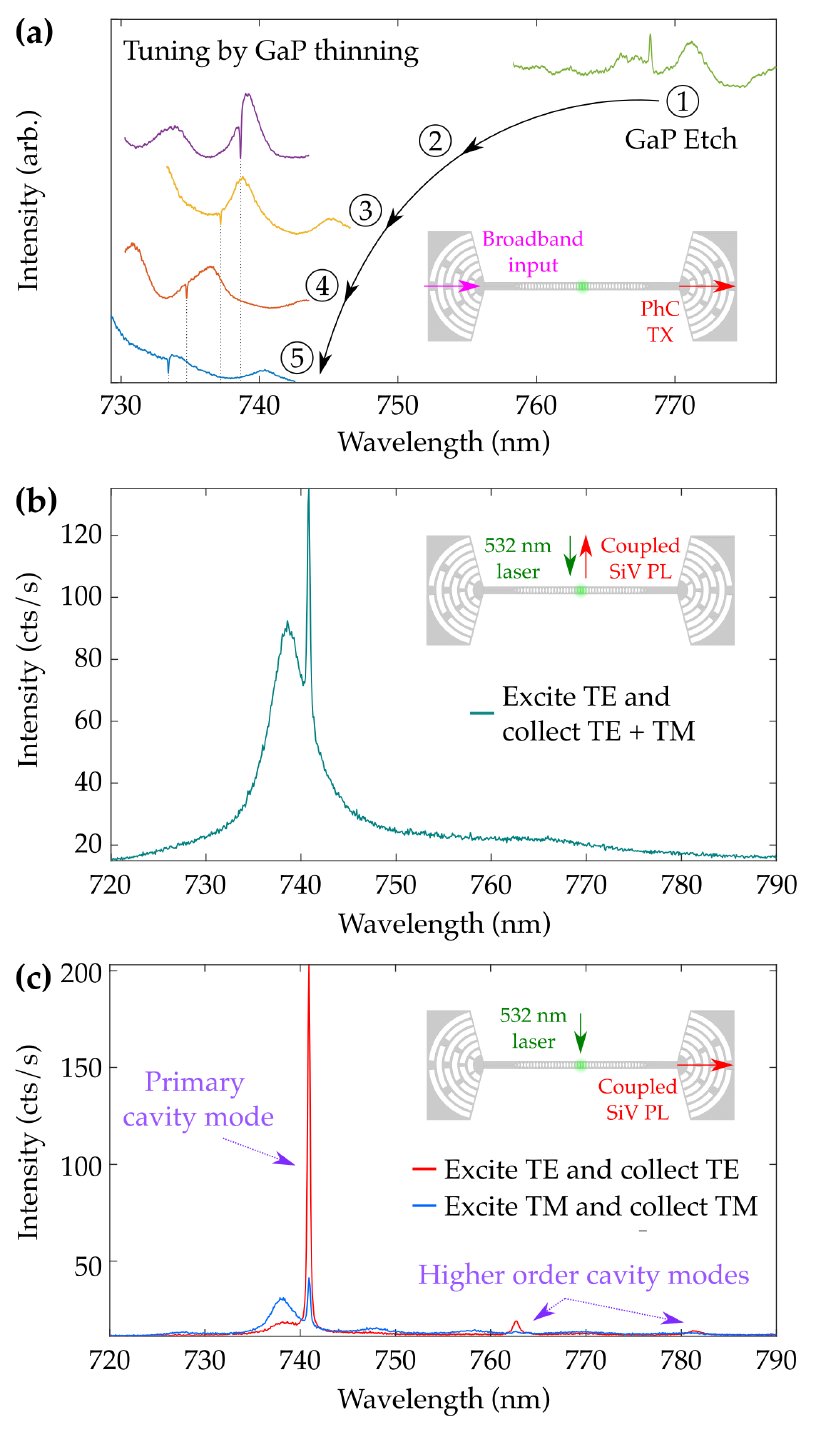}
\caption{\textit{\textbf{(RT measurements)}} (a) The transmission spectra of one specific GaP-on-SiO$_2$ cavity tracked through multiple GaP thinning plasma etch cycles. The resonance is blue shifted from 769\,nm to 733\,nm as the GaP thickness is reduced. (b) Spectra of a GaP-on-diamond cavity with the excitation and PL collection at the center of the cavity. The ensemble SiV PL can be observed with a sharp peak at 741\,nm indicating the cavity mode. (c) Spectra of a GaP-on-diamond cavity with grating coupled PL collection. The cavity mode is significantly brighter and exhibits strong polarization preference for TE collection (matching the designed TE cavity mode).
}
\label{fig:RT_phc}
\end{figure}

The GaP-on-oxide cavity devices are characterized before stamp transfer to diamond. 
The resonances of individual devices are identified by using a supercontinuum laser (600 to 800\,nm, see SI.1) to obtain the transmission spectra. The excitation is linearly polarized, and matched to the TE resonator mode. Scattered excitation light was filtered from the collection path via spatial filtering. We measure cavity resonances near 770\,nm (Fig.~\ref{fig:RT_phc}a-1) strongly red-shifted from design. The red-shift is due to the initial GaP height of 300\,nm which is thicker than the design specification of 222\,nm.

To couple the GaP cavity devices to SiV qubits, the distribution of cavity resonances needs to be centered around the SiV ZPL wavelength of 737\,nm. Here we target 731\,nm as stamp transfer from the SiO$_2$ carrier chip to higher index diamond is expected to red-shift the resonances by $\sim7$\,nm. We blue-shift the cavity resonances by iterative thinning of the GaP layer (Fig.~\ref{fig:RT_phc}a: spectra 1--5) using a short blanket GaP RIE plasma etch (15\,s for step-1, 8\,s for steps-2 to 5; GaP etch rate $\approx$ 100\,nm/min). After each etch cycle, the GaP-on-oxide cavity transmission is measured to track the resonance blue-shift.
After thinning, many cavities exhibit quality factors at the spectrometer limit (Q$_{\mathrm{SiO_2}}{>}30,000$). We measure a final GaP thickness of 225\,nm (SI.2). 

A chemical vapor deposition diamond (Element Six, electronic grade, N~$<$~5\,ppb, B~$<$~1\,ppb) is implanted with Si accelerated to 26\,keV (5e11\,cm$^{-2}$ dose, 7$^\circ$ angle) and vacuum annealed at 1200\,$^\circ$C for 2 hrs. During annealing, SiV centers are formed by vacancy diffusion and recombination with the implanted silicon, yielding a layer of SiV centers 20\,nm from the surface ($\sim$ 50 SiV centers per $\sim$ 800\,nm diameter excitation spot).

To transfer the cavities to the diamond, a stamping procedure~\cite{raha2021telecom,huang2021building} is utilized, as shown in (Fig.~\ref{fig:phc_stamp}:6-10). The process is carried out in a flip-chip wafer-bonder which controls both pressure and temperature.
The process uses a PDMS stamp fabricated into a mesa structure and an intermediate polypropylene carbonate (PPC) film. The PDMS/PPC stack is brought in contact with the devices and heated to 60$^\circ$\,C to promote adhesion. Next, the PDMS/PPC/cavity stack is lifted-off the carrier substrate. The diamond substrate is cleaned in a piranha 3:1 bath prior to stamping. The PDMS/PPC/cavity stack is aligned and pressed down on the diamond substrate. The stamp and diamond are heated to 130$^\circ$\,C, which softens the PPC layer to release the PPC film from the PDMS stamp. Finally, we dissolve the PPC in trichloroethylene and soak the diamond in acetone followed by IPA.

\begin{figure*}[ht]
\centering
\includegraphics[width=\textwidth]{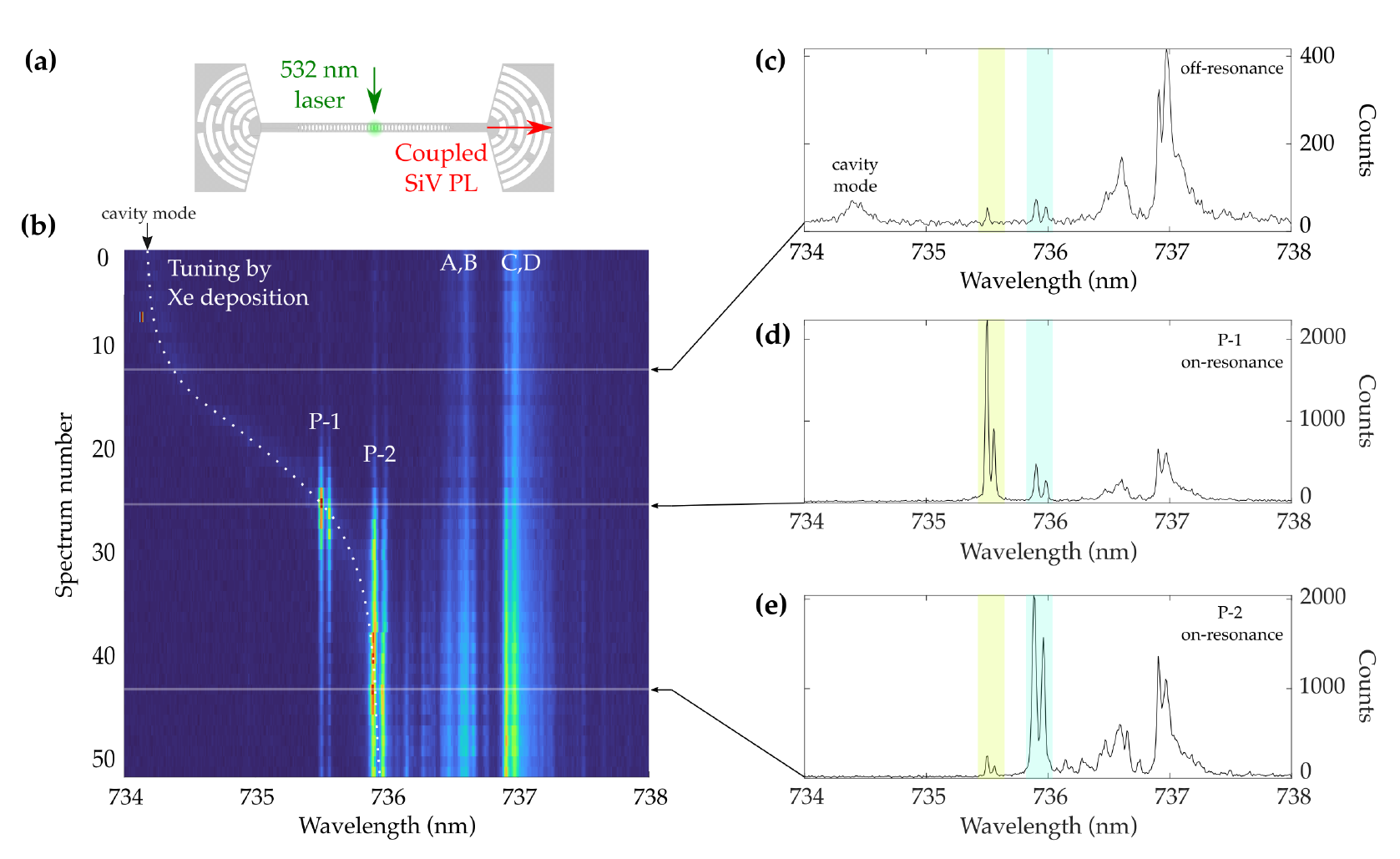}
\caption{\textit{\textbf{(LT 10\,K cavity tuning)}} (a) Illustration of the measurement geometry with the 532\,nm excitation at the cavity and grating coupled SiV PL collection. (b) A sequence of spectra (60\,s exposures) showing the cavity mode red-shifting from xenon gas deposition with a maximum tuning range of $\approx$\,1.9nm. The cavity is tuned through two SiV ZPL pairs (P-1, P-2). (c) Spectra of the cavity off-resonance. (d, e) Spectra of the cavity on-resonance with P-1 (d) and P-2 (e). The green and blue boxes indicate the filtering windows utilized for on-resonance lifetime measurements. 
}
\label{fig:LT_phc}
\end{figure*}

Following the GaP transfer to diamond, cavity transmission measurements are performed. We observe resonances for all 28 intact transferred cavities, with resonance wavelengths in the range of 730 to 755\,nm. Remarkably, many stamped devices show cavity resonances with quality factors between 5,000 to 8,900, close to the design goal of 10,000 (SI.3.).

\begin{figure*}[ht]
\centering
\includegraphics[width=\textwidth]{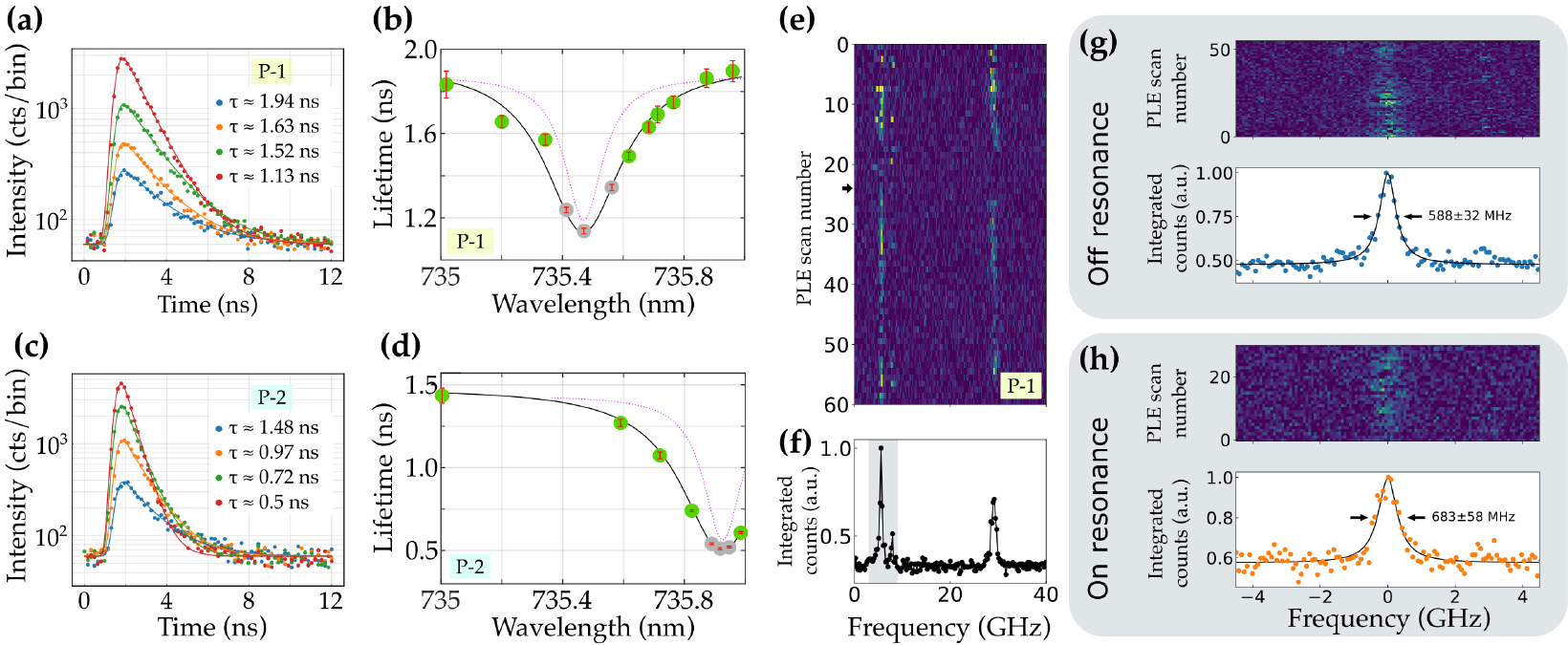}
\caption{\textit{\textbf{(LT 10\,K measurements)}} Time-resolved PL and extracted excited-state lifetime of SiV pairs (a, b) P-1 and (c, d) P-2 as a function of cavity detuning. We observe an enhancement of the decay rate by a factor of $1.7\pm0.1$ and $3.0\pm0.1$ for P-1 and P-2 respectively. The green points in (b, d) correspond to measurement sequences in which the cavity could be resolved in spectra. In instances where the cavity and ZPL significantly overlapped (gray points), the cavity could not be distinguished and so the corresponding points were positioned near the expected detuning of the cavity.
A Lorentzian fit (black) to the lifetime data results in a linewidth that is broader than the cavity linewidth (dotted-pink line).
(e) PLE scans on ZPL P-1 when the cavity is significantly red detuned. Each scan is taken over 2\,s at 10\,s intervals. After each scan, a 532\,nm pulse was occasionally used to reset the SiV charge state if ionization occurred; one such instance is indicated by the black arrow at scan 24. (f) Integrating all scans in (e) shows two bright lines separated by around 25\,GHz approximately corresponding to the spacing observed in spectra. The highlighted line near 5 GHz is further examined by tuning the cavity from (g) off resonance to (h) on resonance while continuously performing PLE scans. The linewidth is observed to broaden by $95\pm66$\,MHz when the cavity is tuned on resonance which is approximately consistent with an enhancement of 1.7 of the decay rate.
}
\label{fig:lifetime}
\end{figure*}

Room-temperature photoluminescence (PL) measurements are performed to study SiV coupling to the stamped GaP photonic devices. We excite with an off-resonant 532\,nm laser and collect SiV PL normal to the sample at the excitation position (Fig.~\ref{fig:RT_phc}b). At room-temperature, the SiV PL spectra consists of a zero-phonon line (ZPL) at 738\,nm (full-width-half-max (FWHM) of 5\,nm) and a broad phonon-sideband (PSB) (744 to 800\,nm). We observe increased SiV ZPL emission at the cavity resonance along with a significant background ZPL contribution from other SiV centers within the excitation spot. By collecting the SiV PL coupled into the waveguide at the output grating, we observe a significantly larger fraction of SiV ZPL intensity at the cavity resonance with minimal background SiV contribution (Fig.~\ref{fig:RT_phc}c). 
Further, we observe a clear polarization dependence for the coupled SiV PL consistent with the designed TE cavity mode. These results confirm that an optical interface has been established between the diamond and the stamped GaP PhC cavity.

When cooled to 10\,K, the cavity mode of the devices are blue-detuned from room-temperature by $\sim$5.5\,nm due to the temperature dependence of the GaP index. Utilizing the 532\,nm off-resonant excitation at the cavity center (Fig.~\ref{fig:LT_phc}a), we can monitor the cavity mode and SiV ZPL transitions at the output grating. Typical spectra shows four bright lines near 737\,nm corresponding to the characteristic A--D transitions for SiV~\cite{hepp2014electronic}, together with a number of faint detuned lines associated with strained single SiV centers. The cavity mode can be red-shifted by injecting xenon gas into the sample chamber. The xenon condenses on the sample, increasing the effective index of the cavity mode and resulting in a controlled resonance shift as demonstrated in Fig.~\ref{fig:LT_phc}b. The maximum range of the xenon-induced red-shift is 1.9\,nm. 
Although all 28 intact devices from the stamped array exhibit cavity resonances, only the nine devices having resonances within 3\,nm of $737$\,nm are utilized for xenon tuning measurements (Fig.~\ref{fig:LT_phc}c). For all nine devices, we are able to observe significant increase in PL on resonance for a range of SiV transitions from blue to red detuned with respect to the ensemble A--D lines (Fig.~\ref{fig:LT_phc}b; other devices in SI.4).

To facilitate coupling to single SiV transitions and for ease of cavity-mode identification, we focus on a device with isolated blue-detuned SiV transitions. We select device-3 with $Q_{\mathrm{Dia}}=$4,100 which exhibits the maximum increase in PL of all devices. For this device, the cavity mode is tuned through two pairs of SiV ZPL denoted P-1 and P-2 (Fig.~\ref{fig:LT_phc}c-e). In order to quantify the SiV-cavity coupling we first perform time-resolved PL measurements while tuning the cavity over P-1 and P-2.
We excite at the cavity with pulsed excitation at 704\,nm (500\,\textmu W average power, 2\,ps pulse width and 80\,MHz repetition rate) and collect PL through the output grating, thus filtering only the cavity-coupled SiV.
We then spectrally filter the PL via a spectrometer (0.2\,nm bandwidth, centered on P-1 or P-2) and detect the PL with an avalanche photodiode. A low-intensity continuous-wave (20\,\textmu W) 532\,nm laser is added to help stabilize the negative SiV charge state (SI.6). 
By fitting the temporally resolved PL (Fig.~\ref{fig:lifetime}a,c, see SI.8 for details), we observe a reduction of the lifetime on resonance by a factor of
\begin{align*}
    \frac{\tau_{\text{off-res}}^{\text{P-1}}}{\tau_{\text{on-res}}^{\text{P-1}}} = 1.7 \pm 0.1, \hspace{2em} \frac{\tau_{\text{off-res}}^{\text{P-2}}}{\tau_{\text{on-res}}^{\text{P-2}}} = 3.0 \pm 0.1.
\end{align*}
A plot of the lifetime as a function of cavity detuning (Fig.~\ref{fig:lifetime}b,d) exhibits a Lorentzian lineshape, with a linewidth that is broader than the cavity resonance (FWHM$\approx$0.17\,nm). This is consistent with the broadening expected due to the contributions from the two SiV transitions comprising each pair (with separation of about $0.06$\,nm and $0.08$\,nm for P-1 and P-2 respectively).   

Next, we perform photoluminescience excitation (PLE) spectroscopy to characterize the optical linewidths off and on cavity resonance.
The SiV is resonantly excited via a low-power tunable diode laser which is scanned over the ZPL transition under consideration while the phonon sideband PL is collected ($\lambda>760$\,nm) from above the cavity (as this emission is not resonant with the cavity).
We observe distinct and consistent lines corresponding to ZPL P-1 (Fig.~\ref{fig:lifetime}e-h) and P-2 (SI.5) by exciting either through the input grating or directly on the cavity.
To maintain constant excitation intensity at the SiV during the scan, we select the latter configuration. Between different SiV centers, the excitation power was varied from 100\,nW to 1\,\textmu W to optimize the signal-to-noise while minimizing SiV ionization. Additionally, a low-intensity blue LED and/or 532\,nm pulse enabled consistent repumping of the SiV (SI.5).

We focus on ZPL P-1 due to the larger tuning range of the cavity.
Fig.~\ref{fig:lifetime}e shows a series of PLE scans over ZPL P-1 with the cavity red detuned by approximately 0.7\,nm.
Two bright lines are observed to be separated by about 25\,GHz, consistent with the separation measured in PL.
The off-resonant lifetime of ZPL P-1 measured in Fig.~\ref{fig:lifetime}a was determined to be $\tau_{\text{off-res}}^{\text{P-1}}=1.94\pm0.06$\,ns which corresponds to a lifetime-limited linewidth $1/2\pi\tau_{\text{off-res}}^{\text{P-1}}= 82 \pm 0.4$\,MHz.
The additional broadening of the lines is understood to be a result of a combination of homogeneous phonon-broadening (limit of $\sim$250\,MHz at 10\,K ~\cite{jahnke2015electron}) and spectral diffusion.
Because the broadening due to pure lifetime reduction is expected to be $ (1/2\pi\tau_{\text{on-res}}^{\text{P-1}}) - (1/2\pi\tau_{\text{off-res}}^{\text{P-1}}) = 60$\,MHz, we focus on the highlighted line near $5$\,GHz as it is significantly narrower than the other line near 30\,GHz. While performing PLE scans centered on this line, we tune the cavity from the original red-detuned position into resonance (Fig.~\ref{fig:lifetime}g,h).
We measure a modest broadening of the overall linewidth by $\Delta\Gamma=95\pm66$\,MHz which is consistent with the estimated broadening via the lifetime measurements.

The lifetime and linewidth measurements can be utilized to determine the Purcell enhancement and cooperativity of the SiV-cavity system. The SiV-cavity system can be modeled as an open system where the cavity is a purely additive decay channel with cavity loss rates $\kappa\equiv \omega/Q$ and defect loss rates $\gamma$ (including both radiative and non-radiative channels outside of the cavity). A Jaynes-Cummings Hamiltonian with single-photon Rabi frequency $g$ describes the SiV-cavity interaction (SI.7). If we consider only contributions to $\gamma$ which describe population decay (i.e. ignore pure dephasing) and assume that the system is in the ``bad-cavity'' regime ($\kappa \gg g, \gamma$), we may define the Purcell enhancement \textit{F} as the multiplicative increase in the decay rate of the desired transition \cite{faraon2011natphoton},
\begin{equation} \label{eq:purcell_enhancement}
    F \equiv 1 + F_P 
    \left(\frac{\va{E} \cdot \va{\mu}}{ |\va{E}_{\text{max}}| |\va{\mu}| } \right)
    \frac{(\kappa/2)^2}{(\kappa/2)^2 + \delta^2} \;.
\end{equation}
Here $\va{E}/|\va{E}_{\text{max}}|$ is the unit-normalized mode electric-field profile, $\delta$ is the SiV-cavity (angular) frequency detuning, and $\va{\mu}/|\va{\mu}|$ is a unit vector in the direction of the transition's dipole moment.
The constant $F_P\equiv (3/4\pi^2)(\lambda/n)^3(Q/V)$ is the well-known Purcell factor for a cavity with mode volume $V$ at wavelength $\lambda/n$ \cite{purcell1946physrev}.

Further assuming that the emitter's proximity to the cavity does not substantially modify $\gamma$, we can determine the ratio of the off-resonant ($\delta\to\infty$) and on-resonant ($\delta=0$) lifetimes $\tau$ to be given by 
\begin{equation} \label{eq:lifetime_ratio}
    \frac{\tau_{\text{off-res}}}{\tau_{\text{on-res}}} = 1 + \frac{4g^2}{\kappa\gamma} \;.
\end{equation}
Comparing Eq.~\ref{eq:lifetime_ratio} to Eq.~\ref{eq:purcell_enhancement} enables estimation of the resonant Purcell enhancement via the relation
\begin{equation}
    F = 1 + \frac{\gamma}{\gamma_{eg}} \left( \frac{\tau_{\text{off-res}}}{\tau_{\text{on-res}}} - 1 \right) \;,
\end{equation}
where $\gamma_{eg}$ is intrinsic spontaneous emission rate of the relevant transition. The factor $\gamma_{eg}/\gamma$ is an intrinsic (cavity-independent) property of the emitter and may be estimated from the quantum efficiency $\eta$, Debye-Waller factor $DW$, and ZPL branching ratio $\xi$ as $\gamma_{eg}/\gamma = \eta \cdot DW \cdot \xi$ (SI.7).
Taking conservative estimates of $\eta = 0.3$ \cite{becker2017pssa}, $DW = 0.7$ \cite{neu2011newjphys}, and $\xi=0.325$ \cite{zhang2018strongly} provides an upper-bound of $(\gamma_{eg}/\gamma)_{\text{max}} \approx 0.126$, and thus a lower bound on the Purcell enhancement $F_{\text{min}}$ of P-1 and P-2. Using the measured lifetime data (Fig.\ref{fig:lifetime}) we find
\begin{align*}
    F_{\text{min}}^{\text{P-1}} \approx 11, \hspace{2em} F_{\text{min}}^{\text{P-2}} \approx 30
\end{align*}
which are within the range of feasible Purcell enhancement (Fig.~\ref{fig:phc_design}(f,g)).

Additionally, the cooperativity $C$ of the SiV-cavity system can be defined as $C \equiv 4g^2/\kappa(\gamma + \gamma_d)$ where $\gamma_d$ is the pure-dephasing rate.
Using the total off-resonant linewidth $(\gamma + \gamma_d)/2\pi = 588\pm32$\,MHz and lifetime-limited linewidth $\gamma/2\pi = 82.0 \pm 0.4$\,MHz, we are able to estimate the cooperativity of the P-1 transition under consideration as
\begin{equation}
    C^{\text{P-1}} = \frac{\gamma}{\gamma + \gamma_d} \left(\frac{\tau_{\text{off-res}}}{\tau_{\text{on-res}}} - 1 \right) = 0.10 \pm 0.01.
\end{equation}
Reducing the temperature to eliminate the thermal photon broadening beyond the lifetime limit is expected to reduce the total linewidth to $(\gamma + \gamma_d)/2\pi \approx 420$\,MHz.
Provided the other broadening mechanisms are temperature independent, this would result in a cooperativity of $C^{\text{P-1}}_{\text{0\,K}} \approx 0.14$.
If the SiV linewidth was lifetime limited (i.e. no pure dephasing), then the maximum expected cooperativity for P-1 is given to be $C_{\text{max}}^{\text{P-1}} = 0.7 \pm 0.1$.
Similarly for P-2, assuming the pure depahsing is of similar magnitude ($\gamma_d/2\pi\approx 506$\,MHz), we estimate a cooperativity of $C^{\text{P-2}}\approx 0.35$, increasing to $C^{\text{P-2}}_{\text{0\,K}}\approx0.45$ below the thermal-broadening limit, and with a maximum value for a lifetime-limited linewidth of $C_{\text{max}}^{\text{P-2}} = 2.0 \pm 0.1$.

Although the SiV studied here do not have lifetime-limited linewidths, similar ranges of broadening were observed in other implanted SiV that were not coupled to devices (SI.5). 
This suggests that a sample with improved defect linewidths could potentially achieve cooperativities $C>1$ with little modification to the fabrication process.
Additionally, since current devices are operating near the design limit, higher $Q$ can be achieved with further design optimization.
Relaxing the design limit on the minimum PhC hole dimension from $h_x=70$\,nm to $60$\,nm leads to an increase of $Q$ by a factor of $5$. Additionally, stamping of the cavity along the $\langle110\rangle$ crystal axis or utilizing a diamond with $\langle111\rangle$ surface orientation could result in improvements to the coupling coefficient by up to a factor of 4. This indicates that $C\gtrsim10$ can be achieved in hybrid device geometries despite the low refractive index contrast. 
This regime is promising for applications such as optical spin readout~\cite{rogers2014all}, quantum-optical switches and tunable single-photon sources \cite{sipahigil2016integrated}, and quantum repeaters \cite{dur1999repeaters}. 

In summary, we have demonstrated significant coupling between a GaP cavity and diamond SiV qubit avoiding all fabrication of the diamond substrate. Before integration the cavities can be characterized and tuned close to the specific target resonance wavelength. This enables optimization of the SiV-cavity coupling yield with the potential for minimal disruption to the qubit environment within the diamond. For the transferred GaP-on-diamond cavities, we measure quality factors close to the design target. Current device properties should enable cooperativity $C>1$, with a promising outlook for $C\sim10$ with modest design improvements. These results highlight the promise of hybrid photonic architectures for quantum network applications, even in relatively low-index contrast material combinations. 

\section{Supporting information}
Schematic of microscope system, Images of fabricated devices, Room-temperature transmission spectra of GaP-on-diamond devices, Tuning curves on different devices, Resonant excitation measurements on both device and non-device SiV, Intensity dependence of SiV photoluminescence with 532\,nm and 700\,nm excitation, Theory and simulation of the SiV-cavity system, Fits and error approximation.

\section{Acknowledgements}
This material is based upon work supported by Department of Energy, Office of Science, National Quantum Information Science Research Centers, Co-design Center for Quantum Advantage (C2QA) under contract number DE-SC0012704 and National Science Foundation Grant No. ECCS-1807566. N.S.Y.\ was supported by the National Science Foundation Graduate Research Fellowship Program under Grant No.~DGE-2140004.  A.A.\ was supported by a Post Graduate Scholarship from the Natural Sciences and Engineering Research Council of Canada (PGSD3 - 545932 - 2020). D.H.\ was supported by the National Science Scholarship from A*STAR, Singapore. The photonic devices were fabricated at the Washington Nanofabrication Facility, a National Nanotechnology Coordinated Infrastructure (NNCI) site at the University of Washington which is supported in part by funds from the National Science Foundation (awards NNCI-2025489, 1542101, 1337840 and 0335765). The authors acknowledge the use of Princeton’s Imaging and Analysis Center, which is partially supported through the Princeton Center for Complex Materials (PCCM), a National Science Foundation (NSF)-MRSEC program (DMR-2011750) as well as the Princeton Micro-Nano Fabrication Lab.

\bibliography{main}
\end{document}


\title{Supplementary Information: Hybrid Integration of GaP Photonic Crystal Cavities with Silicon-Vacancy Centers in Diamond by Stamp-Transfer}

\author{Srivatsa Chakravarthi}
\email{srivatsa@uw.edu}
\affiliation{University of Washington, Physics Department, Seattle, WA, 98105, USA}
\author{Nicholas S. Yama}
\email{nsyama@uw.edu}
\affiliation{University of Washington, Electrical and Computer Engineering Department, Seattle, WA, 98105, USA}
\author{Alex Abulnaga}
\email{alex.abulnaga@princeton.edu}
\affiliation{Princeton University, Electrical and Computer Engineering Department, New Jersey 08544, USA}
\author{Ding Huang}
\affiliation{Princeton University, Electrical and Computer Engineering Department, New Jersey 08544, USA}
\author{Christian Pederson}
\affiliation{University of Washington, Physics Department, Seattle, WA, 98105, USA}
\author{Karine Hestroffer}
\affiliation{Department of Physics, Humboldt-Universitat zu Berlin, Newtonstrasse, Berlin, Germany}
\author{Fariba Hatami}
\affiliation{Department of Physics, Humboldt-Universitat zu Berlin, Newtonstrasse, Berlin, Germany}
\author{Nathalie P. de Leon} 
\affiliation{Princeton University, Electrical and Computer Engineering Department, New Jersey 08544, USA}
\author{Kai-Mei C. Fu}
\affiliation{University of Washington, Electrical and Computer Engineering Department, Seattle, WA, 98105, USA}
\affiliation{University of Washington, Physics Department, Seattle, WA, 98105, USA}
\affiliation{Physical Sciences Division, Pacific Northwest National Laboratory, Richland, Washington 99352, USA}

\date{\today}

\maketitle


\newpage
\section{Optical measurements}

\begin{figure*}[ht]
\centering
\includegraphics[width=5in]{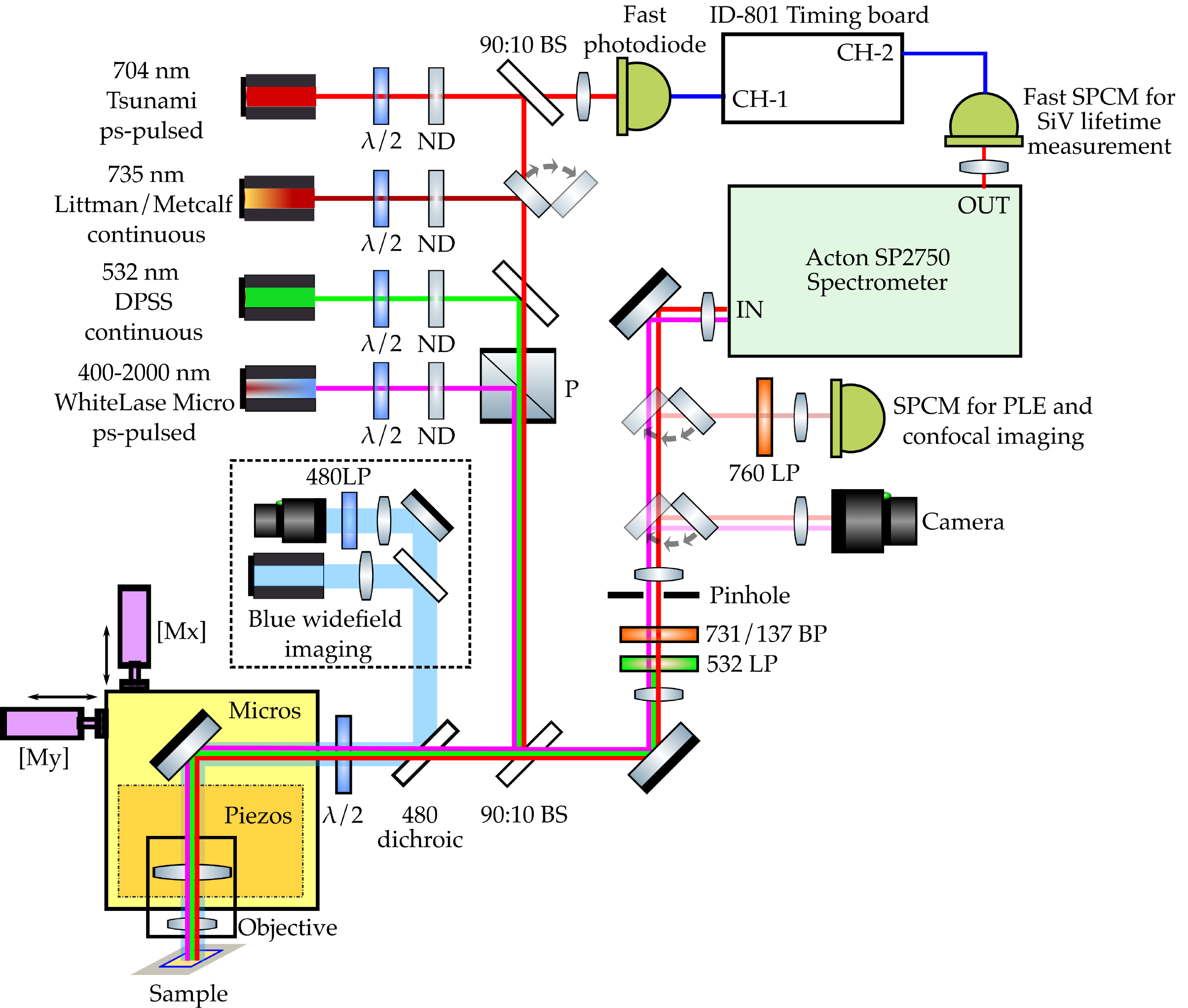}
\caption{Schematic of the optical measurement setup. The GaP-on-diamond devices are placed inside a 10\,K closed cycle He cryostat.}
\label{fig:SI_1}
\end{figure*}

\newpage
\section{Fabricated devices}

\begin{figure*}[ht]
\centering
\includegraphics[width=\textwidth]{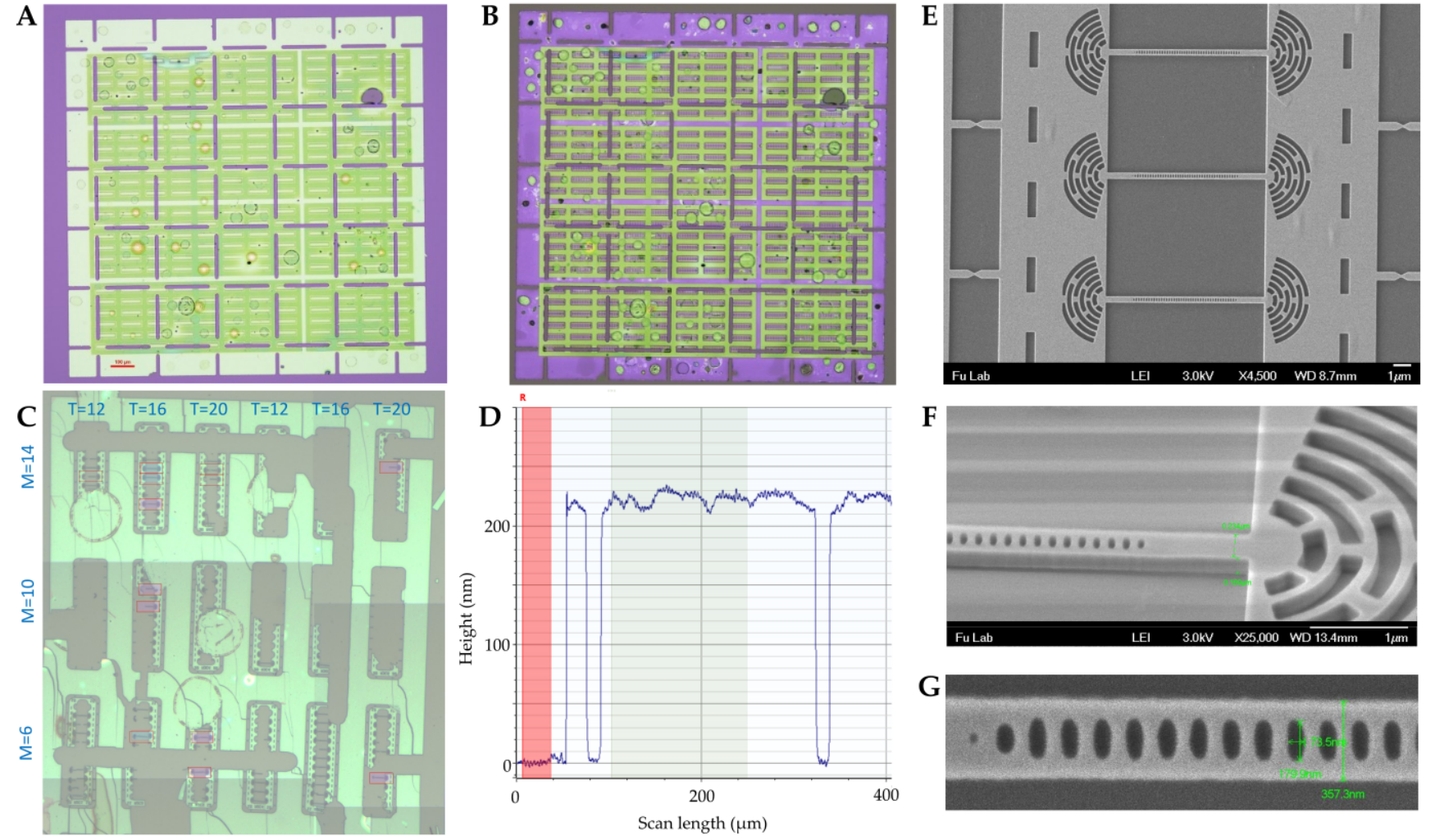}
\caption{(a) A 1.5\,mm$\times$1.5\,mm GaP membrane (300\,nm thick) transferred to a silicon oxide carrier substrate. The photonic devices are aligned to the membrane and patterned with e-beam lithography using HSQ negative-resist. (b) After etching the GaP using (Ar/Cl$_2$/N$_2$) ICP-RIE plasma processing. (c) Devices after stamp transfer to diamond (after pre-transfer characterization and GaP thinning for fine tuning cavity resonances). The stamping yield is low, however many high-Q devices are measured (indicated by the colored boxes). (d) A profilometer scan across the transferred devices showing the average final GaP thickness of 225\,nm. (e) SEM image of an array of PhC devices. (f) An angled SEM image of the PhC and grating coupler. (g) A close up SEM of the PhC holes. The final dimensions of the PhC structure are within 5\,nm of the optimized design.}
\label{fig:SI_2}
\end{figure*}

\newpage
\section{GaP-on-diamond PhC transmission}

\begin{figure*}[ht]
\centering
\includegraphics[width=\textwidth]{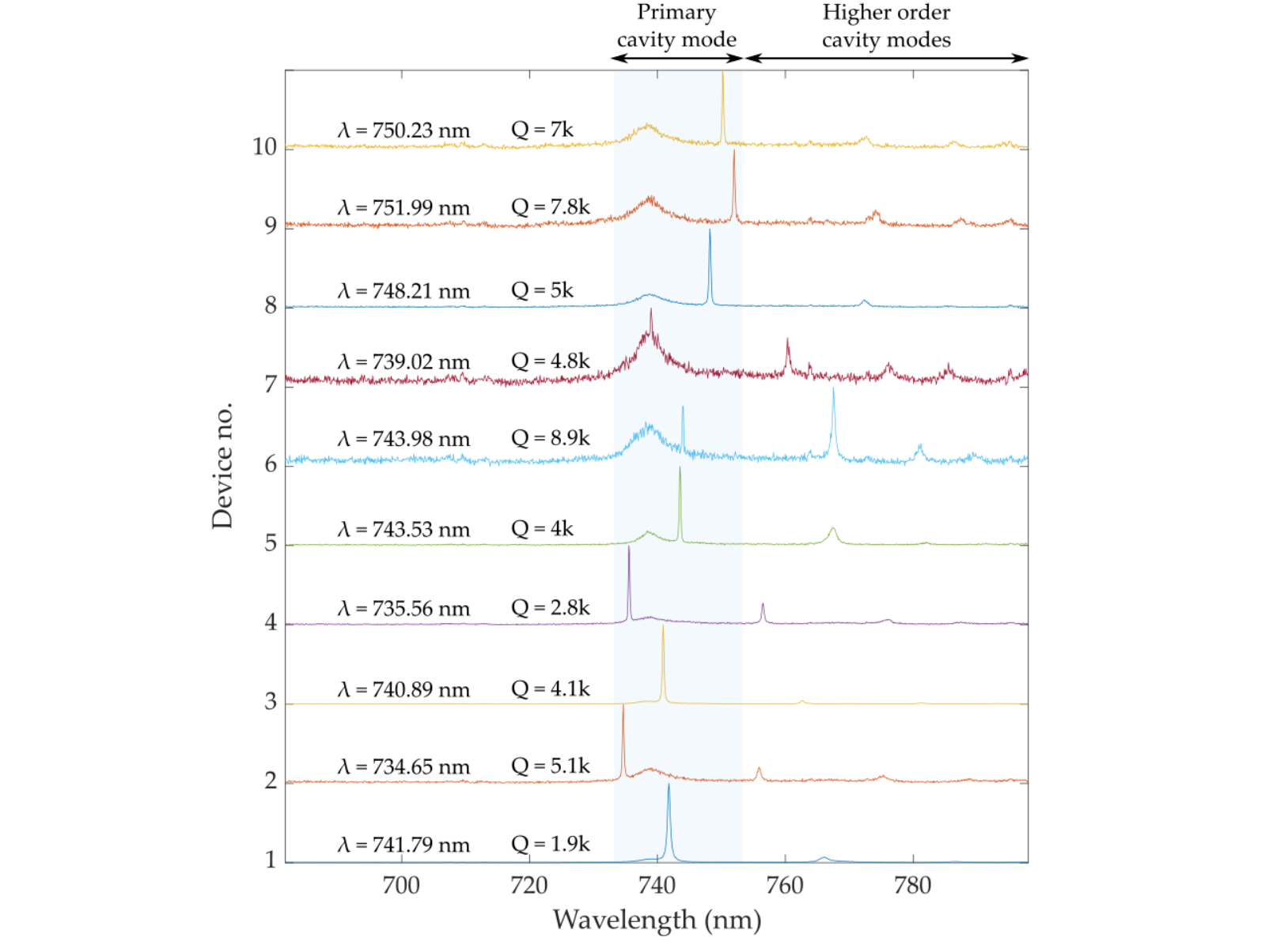}
\caption{Measured cavity resonances of ten GaP-on-diamond PhC devices at room temperature. The devices are excited at the cavity with 500\,\textmu W, 532\,nm CW laser and the coupled SiV PL is measured at the output grating on a spectrometer. The resonance center wavelength and $Q$ factor are extracted with a Lorentzian fit. Single SiV centers coupled to Device-3 are discussed in the main text. The cavity resonances blue-shift by 5.5\,nm after cooling down to 10\,K.}
\label{fig:SI_3}
\end{figure*}

\newpage
\section{Measured PhC Xenon tuning curves}
\begin{figure*}[ht]
\centering
\includegraphics[width=\textwidth]{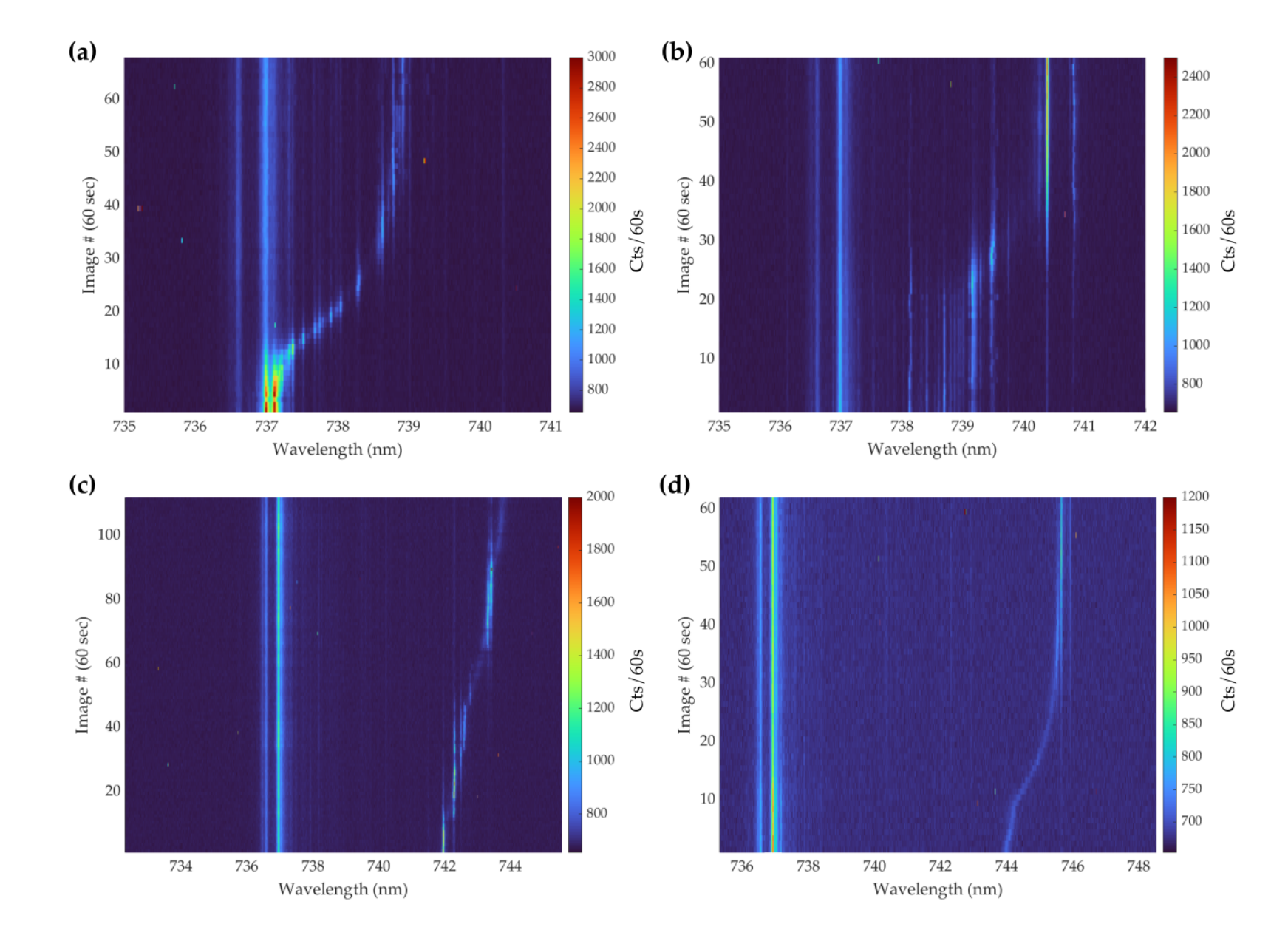}
\caption{(a--d) Show a sequence of spectra captured during the Xe gas deposition process (at 10\,K) for four cavity devices with resonances from 737\,nm (a) to 744\,nm (d). Coupled SiV transitions can be observed throughout the tuning range as the cavity resonances are gradually red-shifted.}
\label{fig:SI_5}
\end{figure*}

\newpage
\section{Resonant excitation of SiV centers}
Photoluminescence excitation (PLE) measurements of SiV centers (both device-coupled and non-device-coupled) were taken with a continuous-wave scanning laser (Sacher Lasertechnik).
We spectrally filter the SiV phonon-sideband using a combination of 760-nm long-pass and 785/62-nm band-pass filters and collect the photoluminescence on an avalanche photodiode (Excelitas).
The output data stream of the detector is temporally correlated with the laser scan signal and an additional analog power meter output which samples the laser.
This power meter monitors the intrinsic variation of power over the scan range and can signify mode hopping of the laser during individual scans as identified by large discontinuous jumps in laser power.
For all PLE data shown in this work, the laser scan range is chosen to avoid significant variation of the power and mode hops during the course of the scan and so this calibration data is not used explicitly.

\subsection{Device-coupled SiV centers}
\begin{figure*}[ht]
\centering
\includegraphics[width=0.6\textwidth]{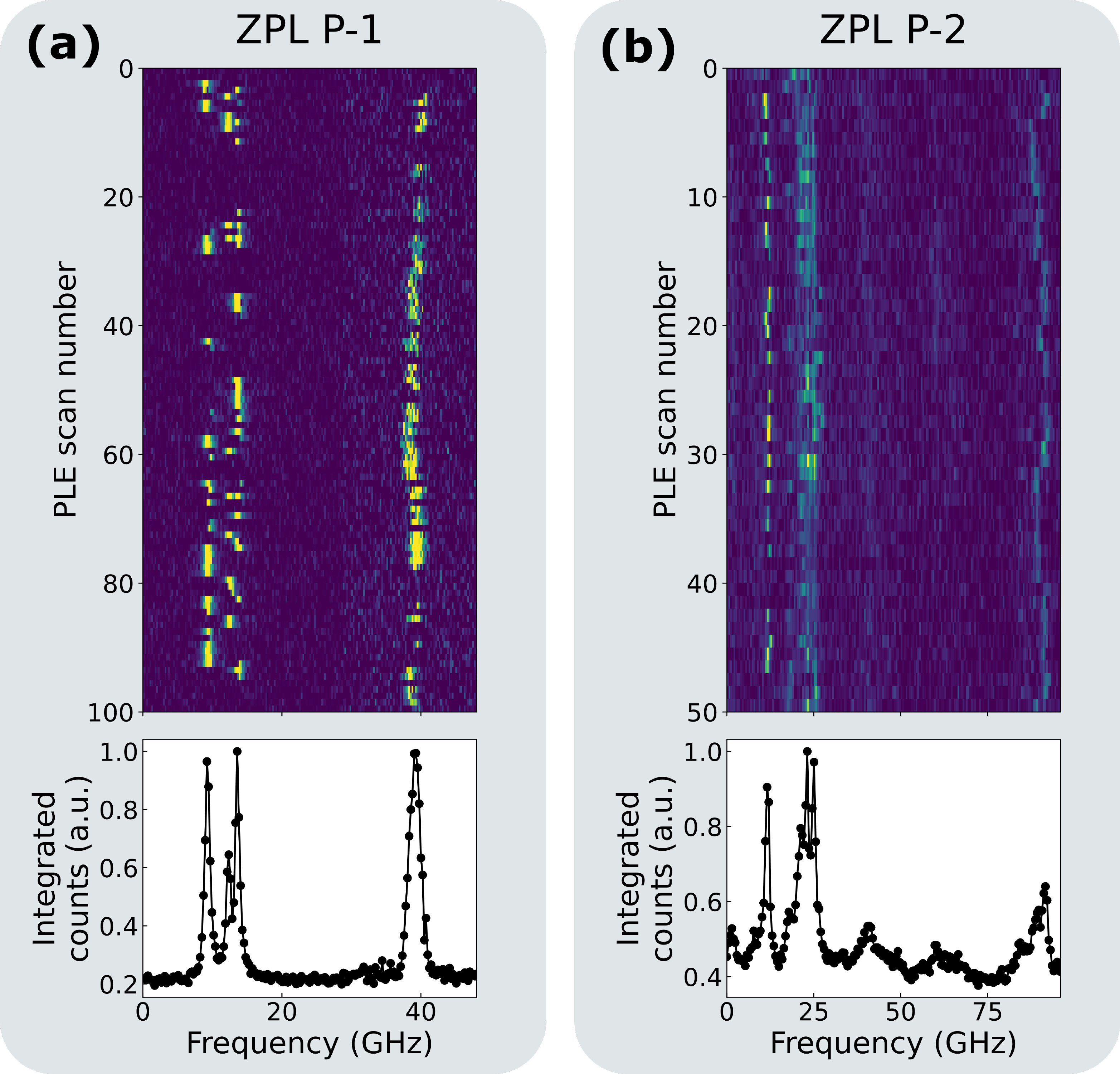}
\caption{\textbf{\textit{(PLE measurements of device-coupled SiV)}} Resonant excitation scans of (a) P-1 and (b) P-2 taken with excitation through the grating and collection from above the cavity.
The integrated counts for the entire scan are shown below, however there is significant spectral diffusion between subsequent scans.
The structure of P-1 matches that of Fig. 5e,f in the main text indicating that the same SiV lines could be identified despite exciting from the top of the cavity.
In both cases, the PLE data shows that there are actually several lines in each set.
Specifically in the case of P-2 it is difficult to distinguish which lines contribute to the measured PL.
}
\label{fig:device_ple_1}
\end{figure*}

As mentioned in the main text, PLE measurements taken on the devices can be done with either excitation through the waveguide or from the top of the cavity.
Exciting through the waveguide is ideal to spatially filter only SiV with some degree of cavity coupling.
PLE scans on the device-coupled SiV (P-1 and P-2) taken while the cavity was on resonance are shown in Fig. \ref{fig:device_ple_1}.

\begin{figure*}[ht]
\centering
\includegraphics[width=0.8\textwidth]{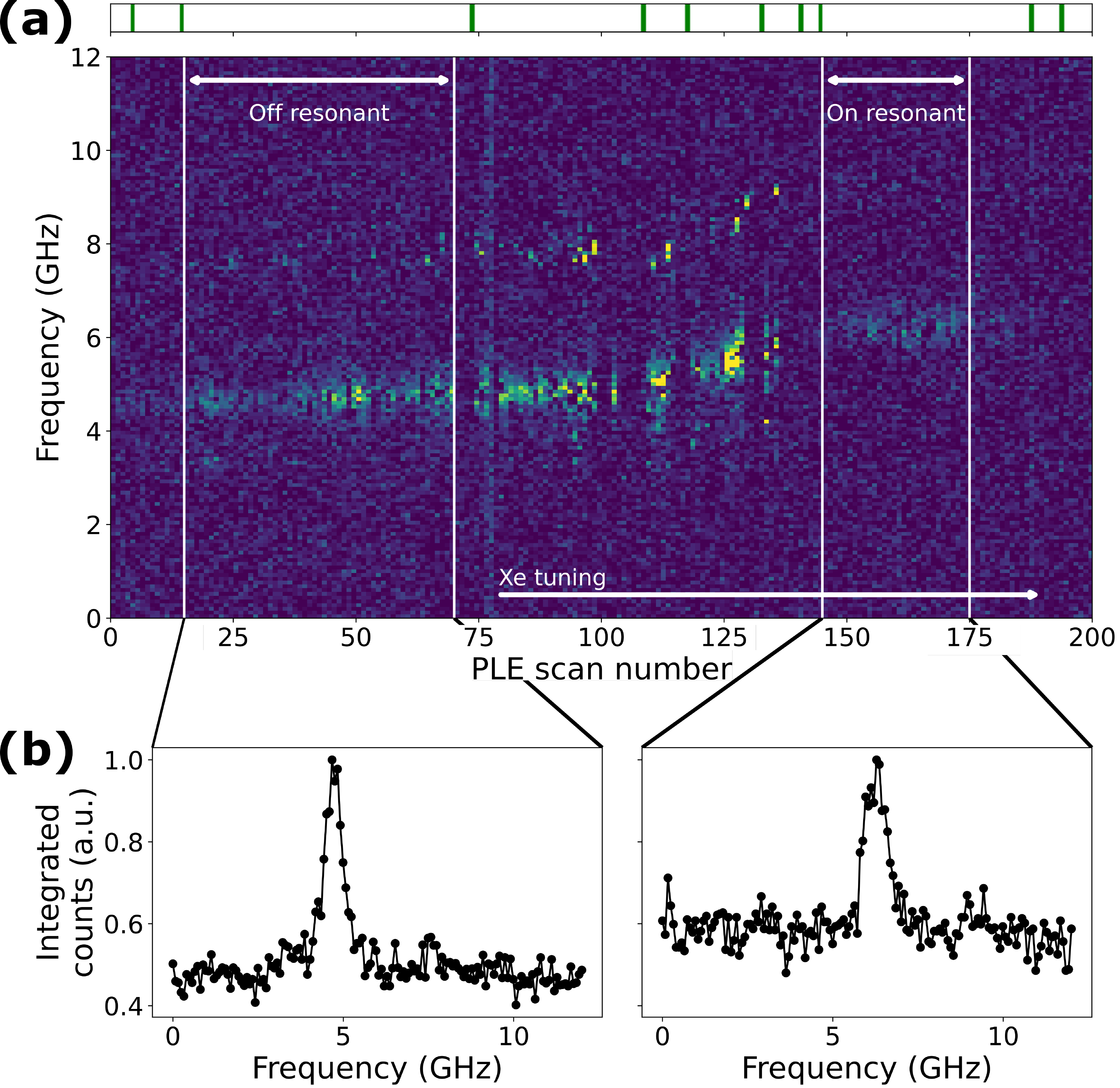}
\caption{\textbf{\textit{(Continuous PLE measurement on ZPL P-1 while tuning the cavity)}} (a) The full series of scans taken while the cavity is tuned on to resonance with P-1 as described in the main text.
The green blocks in the bar above the plot indicates scans after which the green laser was pulsed to repump the charge state.
In scans where the bar is white there was no repump.
The indicated intervals of scans show the regions which were integrated to generate Fig. 5g,h in the main text.
Cavity tuning occurs from the indicated scan (scan \#78) with the preceding scans having larger background due to usage of a flashlight in the room while opening the Xe gas line.
The drift in frequency with additional Xe tuning is of unknown origin.
(b) Replicates the data binning as shown in Fig. 5g,h of the main text.
}
\label{fig:device_ple_2}
\end{figure*}

Because of the well-resolved transitions of P-1, we performed PLE while continuously tuning the cavity to observe a change in the linewidth.
The entire PLE measurement is shown in Fig. \ref{fig:device_ple_2}.
These scans were taken over two seconds each at 10-second intervals.
Over the course of the measurement there were a number of instances where SiV line disappeared which was attributed to ionization.
A green laser pulse was manually triggered with the software after scans denoted with the green marker in order to repump the SiV charge state.
The indicated intervals correspond to the integration regions of Fig.~5g,h in the main text.
At the end of the sequence the SiV line disappeared which is attributed to the degradation in imaging quality as a result of xenon gas condensation.
The tuning rate in this experiment was significantly faster than the the rate in the PL.
At the end of the tuning, resonance with P-1 was confirmed via spectra.

\subsection{Non-device SiV centers}
Additional PLE measurements were taken on implanted SiV that were located around 10 micrometers away from any transferred GaP.
Due to the relatively high density of SiV in the sample, it was generally possible to locate sharp SiV PLE lines within 10\,GHz detuning from the main ensemble.
Additionally, single lines of highly strained SiV with significant detuning could be located spectrally and subsequently scanned over in PLE.
As discussed in the main text, the excitation power which maximized the signal while also minimizing the ionization was generally variable between SiV, ranging from 10--100\,\textmu W.

\begin{figure*}[t]
\centering
\includegraphics[width=0.8\textwidth]{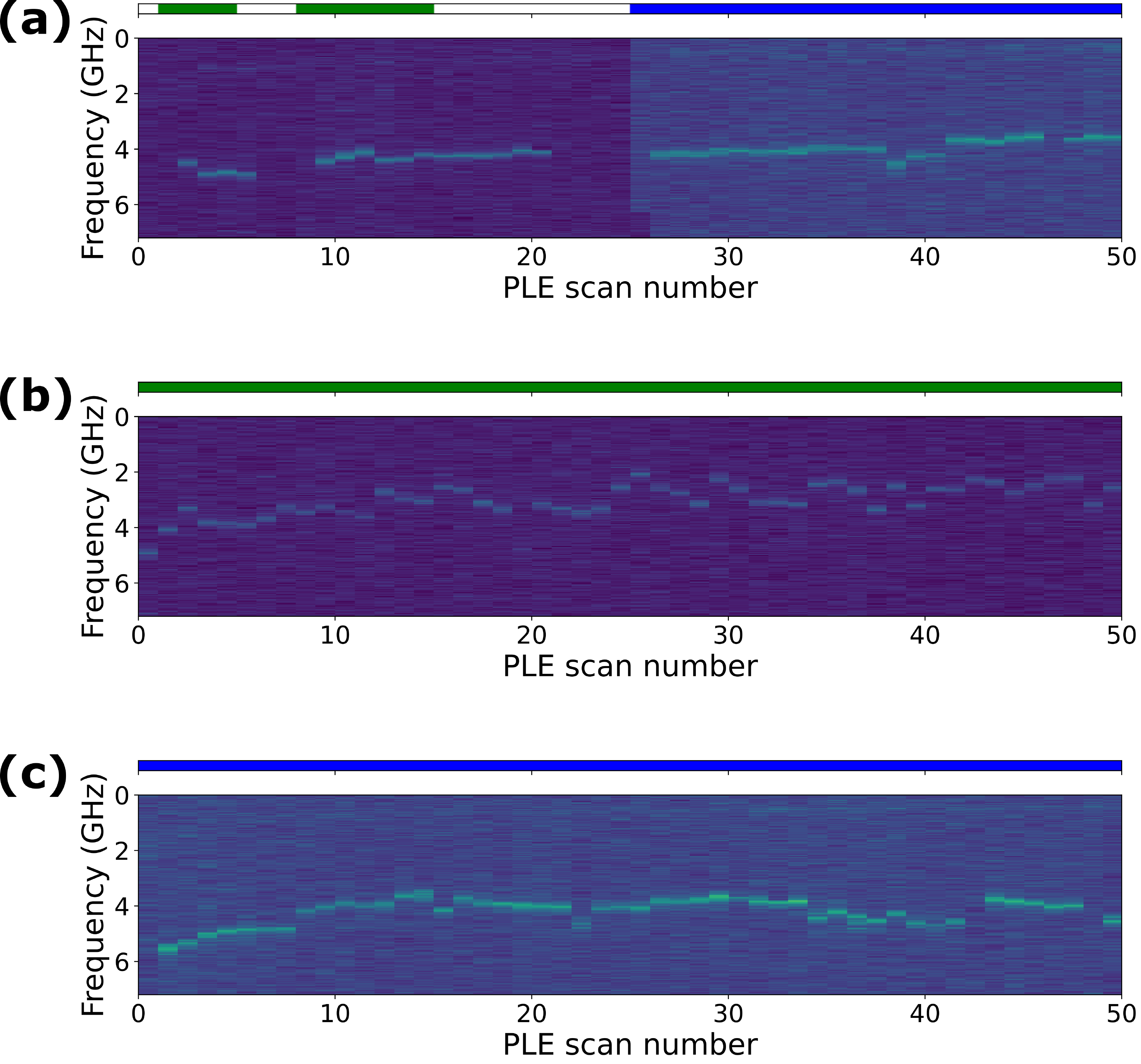}
\caption{\textbf{\textit{(Charge state repumping mechanisms)}} PLE measurements taken on a blue-detuned SiV (near 736.07\,nm) using different repump mechanisms. As before, the bar above the PLE scan figure indicates the scans in which a particular repump method was used.
Here the green color indicates a green repump pulse was performed after that particular scan whereas the blue color indicates a continuous blue LED which is on throughout the indicated scans.
(a) Demonstration of both methods being able to repump the SiV charge state which had gone dark for a few scans.
(b) PLE scans taken with the green pulse occuring after each scan.
(c) PLE scans with the blue LED on for the entirety of the measurement.
The usage of either repump method repeatedly results in a noticable increase of spectral diffusion between subsequent scans.
}
\label{fig:bulk_ple_1}
\end{figure*}

In general, the SiV charge-state stability varied between different SiV and often would be unstable for individual SiV.
We utilized a blue LED ($\sim200$\,nW) and/or green laser ($\sim200$\,\textmu W) to repump the charge state during measurements.
The blue LED is uncollimated (used for wide-field imaging) and so the actual power delivered to the SiV is significantly less than $200$\,nW.
Fig.~\ref{fig:bulk_ple_1} shows a number of different scans on a blue-detuned SiV located near 736.07\,nm utilizing different repump mechanisms.
It was consistently observed that both the continuous blue LED and interleaved green laser pulse repump schemes resulted in a significant increase to the spectral diffusion between scans relative to the intervals without any repump mechanism.
Despite the increase in spectral diffusion, the linewidth in any particular scan did not appear to change significantly when a repump mechanism was used.

\begin{figure*}[ht]
\centering
\includegraphics[width=0.9\textwidth]{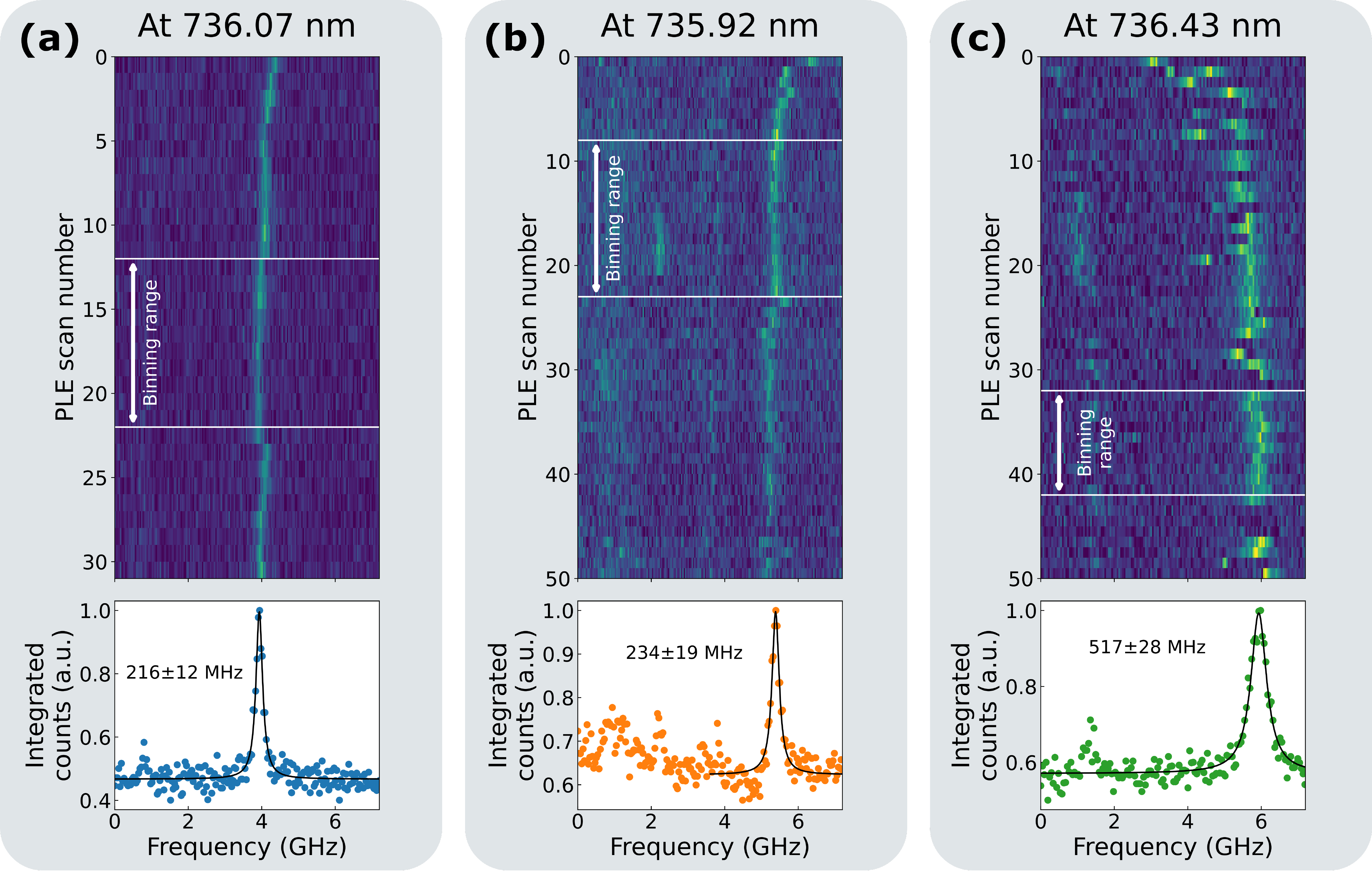}
\caption{\textbf{\textit{(Linewidths of implanted non-device-coupled SiV)}} PLE scans over three different SiV transitions in bulk diamond (a) without and (b,c) with a continuous blue repump.
The indicated binning ranges are integrated and a Lorentzian fit is taken to determine the linewidth of the transition.
For (b), the fit was taken over the portion of data for which the curve is drawn in order to avoid noise from the additional peaks.
The linewidths of the SiV vary from near the thermally broadened limit of 250\,MHz for scans (a) and (b) to more than 500\,MHz in (c).
}
\label{fig:bulk_ple_2}
\end{figure*}

Finally, Fig.~\ref{fig:bulk_ple_2} shows different sets of PLE scans taken on three different SiV with varying degrees of detuning.
We show here two cases of SiV with linewidths near the thermally broadened limit of $250$\,MHz (at 10\,K) and one case in which the SiV had a significantly broadened linwidth of around $500$\,MHz.
A range of spectral intensity, stability, and linewidths can be observed.

\newpage
\section{Intensity dependence of SiV with 532 nm and 700 nm excitation}

\begin{figure*}[ht]
\centering
\includegraphics[width=\textwidth]{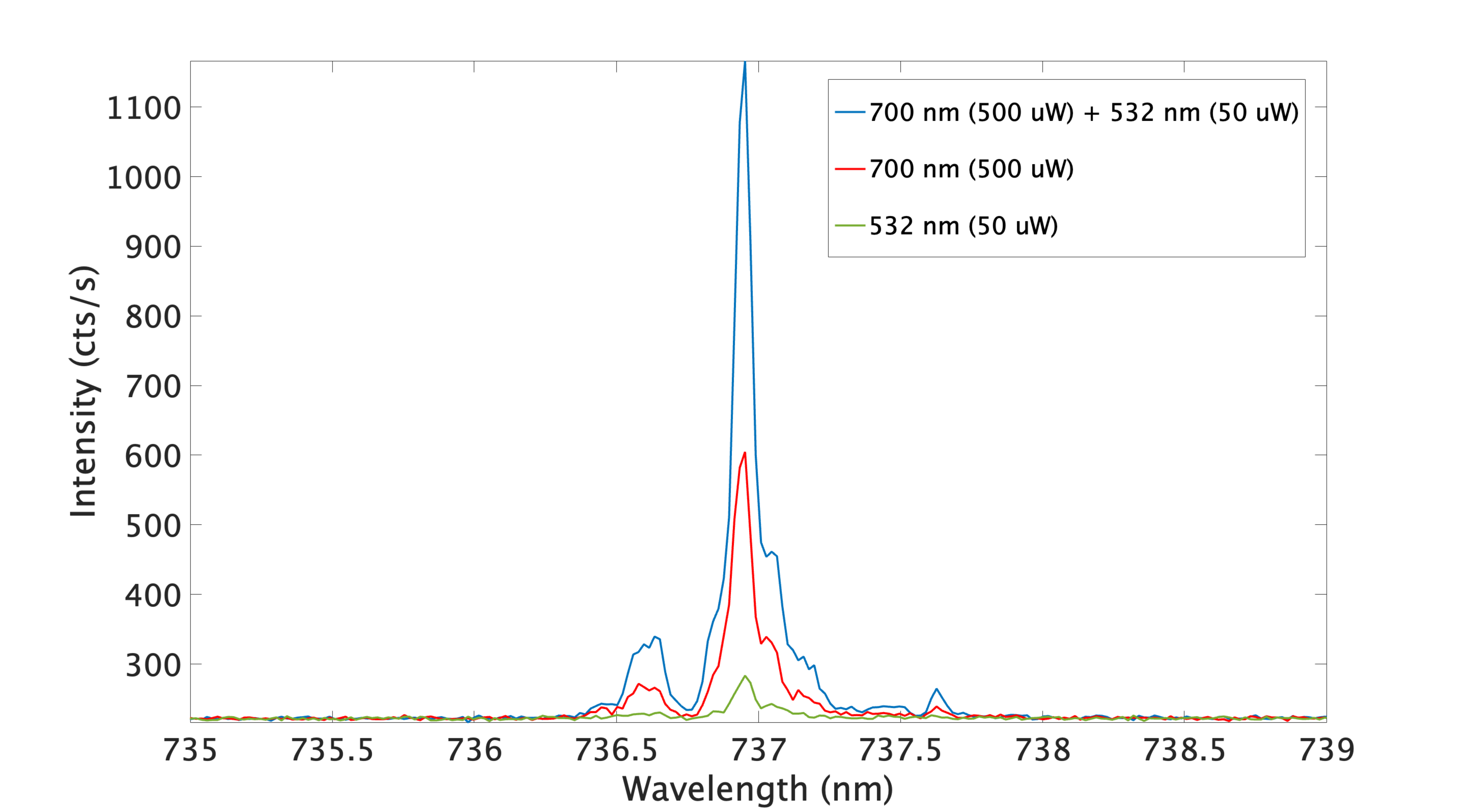}
\caption{Variation of SiV photo-luminescence intensity under 532\,nm and 700\,nm excitation separately and simultaneously.  When excited simultaneously, the observed intensity is more than the sum of the intensities from each excitation separately. The 700\,nm excitation comes from a pulsed Tsunami laser while the 532\,nm excitation is the output of a continuous diode laser.}
\label{fig:SI_6}
\end{figure*}

Optically exciting defects at different wavelengths can reveal complex behavior resulting from that defect's excited state structure and charge exchange with nearby charge traps. Here we excite with 700\,nm excitation, which excites SiV from the ground state into the phonon side-band of the optically-active excited state. 532\,nm excitation excites the SiV from the ground state into a different, higher lying excited state \cite{haussler}. We observe that the SiV exhibit a non-linear increase in PL intensity when both lasers are on simultaneously, a clear signature of laser-induced deshelving of at least one meta-stable state. We suspect that the metastable states are likely the alternate charge states SiV$^{0}$, and SiV$^{2-}$ formed either from direct laser ionization or charge capture from nearby charge traps \cite{jayakumar_long-term_2020,dhomkar_-demand_2018}. Further understanding and engineering of the diamond surface, and SiV photo-dynamics should further stabilize and increase the SiV$^-$ PL intensity.

\newpage
\section{Modeling the PhC-SiV system}
\label{sec:model}
We follow the standard approach \cite{lukin2016notes} to model the SiV-cavity system as a two-level system with energy $\hbar\omega_a$ coupled to a cavity with resonance at $\omega_c$.
The Hamiltonian of the system is given by the Jaynes-Cummings Hamiltonian, here written in the interaction picture with constant-coefficients,
\begin{equation}
    H_{\mathrm{JC}} = \hbar\delta a^\dagger a 
    -\hbar g \big( a^\dagger \dyad{g}{e}
    + a \dyad{e}{g} \big)
\end{equation}
where we define $\delta \equiv \omega_c - \omega_e$ as the detuning and $g$ as the single-photon Rabi frequency which characterizes the coupling strength.

We model dissipation from the SiV-cavity system via the open system formalism, taking the appropriate reservoir assumptions such that the evolution of the SiV-cavity system density operator $\rho$ is given by the Lindblad mater equation (at zero temperature)
\begin{equation}
    \dv{\rho}{t} = \frac{1}{i\hbar} [H_{JC},\rho]
    + \sum_j \Gamma_j \left( 
    L_j \rho L_j^\dagger
    - \frac{1}{2} \{L_j^\dagger L_j, \rho\}
    \right)
\end{equation}
where $[\cdot,\cdot]$ ($\{\cdot,\cdot\}$) is the (anti-)commutator and $\Gamma_j$ is the rate of (population) dissipation via channel $j$ associated to the Lindblad operator $L_j$ defined on the SiV-cavity system Hilbert space.
For the system of interest the dissipation is described by tuples $(\Gamma_j, L_j)$ with
\begin{align*}
    (\gamma, \dyad{g}{e}) &\equiv \text{SiV \textit{decay} outside of the cavity (radiative and non-radiative),} \\
    (\gamma_d, \dyad{e}{e}) &\equiv \text{Pure dephasing of the SiV, and } \\
    (\kappa, a) &\equiv \text{Cavity loss (both radiation and coupling).}
\end{align*}
Note that these rates are given in units of angular frequency.

To model the system analytically we use the so-called stochastic wavefunction formalism.
Our aim is to describe the ensemble-average behavior and so we neglect quantum jumps and model the system using the effective Hamiltonian
\begin{equation}
    H_{\text{eff}} = H_{\text{JC}} 
    - i\hbar \sum_j \frac{\Gamma_j}{2} L_j^\dagger L_j.
\end{equation}
This allows us to describe the dynamics of the system state $\ket{\psi}$ on a \textit{subspace} of the SiV-cavity system Hilbert space provided that all dissipative processes result in projections outside of the subspace under consideration (i.e. dissipation must not occur \textit{within} the subspace).
A necessary constraint, then, is that we must neglect pure dephasing $(\gamma_d, \dyad{e}{e})$ for this analysis.

\subsection{Modification of emitter lifetime}
Our aim is to describe the lifetime measurements (off-resonant, pulsed driving of the SiV).
In this case the decoherence due to pure dephasing can be ignored and the relevant subspace is spanned by the states $\{ \ket{e,0}, \ket{g,1} \}$ where $\ket{x,n}$ corresponds to the atom in state $x$ and $n$ photons in the cavity.
As required in the stochastic wavefunction formalism, the dissipation processes of interest, $(\gamma, \dyad{g}{e})$ and $(\kappa, a)$, result in projections into the state $\ket{g,0}$ which is outside of this subspace.

We consider the system in the so-called leaky-cavity regime distinguished by rates $\kappa \gg g, \gamma$, which is generally true of cavities with moderate to low quality factor $Q = \omega/\kappa$.
The effective Hamiltonian is given by
\begin{equation}
    H_{\text{eff}} = H_{\text{JC}} 
    - i\hbar\frac{\gamma}{2} \dyad{e}
    - i\hbar\frac{\kappa}{2} a^\dagger a.
\end{equation}
The Schr\"odinger equation for a general state $\ket{\psi} = c_e\ket{e,0} + c_g \ket{g,1}$ yields the system of equations
\begin{align}
    \dv{c_g}{t} &= - \left(\frac{\kappa}{2}+i\delta\right) c_g
    - ig c_e , \\
    \dv{c_e}{t} &= -igc_g - \frac{\gamma}{2} c_e.
\end{align}
In the leaky-cavity regime, cavity losses dominate so the state $\ket{g,1}$ dissipates faster than the dynamics of other processes.
As a result, $c_g$ ``reaches equilibrium'' on timescales relevant to the coupling $g$ and SiV emission $\gamma$ rates so we may set $\dv*{c_g}{t} = 0$.
This process is known as adiabatic elimination \cite{lukin2016notes}.

The system of equations can then be solved with initial condition $\ket{\psi(t=0} = \ket{e,0}$ to find the population dynamics of the excited state as $|c_e(t)|^2 = \exp(-\gamma_{\text{eff}} t)$ with
\begin{equation} \label{eq:gamma_eff}
    \gamma_{\text{eff}} = \gamma 
    + \frac{4g^2}{\kappa} \frac{(\kappa/2)^2}{(\kappa/2)^2 + \delta^2}.
\end{equation}
We may compare this effective decay rate to the intrinsic decay rate of the SiV dipole transition given by the Wigner-Weisskopf theory,
\begin{equation} \label{eq:wigner_weisskopf}
    \gamma_{eg} \equiv \frac{\omega_e^3 n |\mu|^2}{3\pi\epsilon_0 \hbar c^2},
\end{equation}
where $|\mu|$ is the amplitude of the dipole moment of the relative SiV transition, $\epsilon_0$ is the permittivity of free space, $c$ is the speed of light, and $n$ is the refractive index of diamond.
For an ideal emitter, when the cavity is \textit{open} such that $\gamma \approx \gamma_{eg}$ (i.e. the cavity does not suppress the emission outside of the cavity, which is the case for evanescently coupled emitters), Eq. \ref{eq:gamma_eff} shows that the cavity coupling $g$ results in an enhanced decay rate of the excited state.
This is the well-known Purcell effect \cite{purcell1946physrev}.

\subsection{Purcell enhancement}
The \textit{Purcell enhancement} is a figure of merit that quantifies the modification of the spontaneous decay rate of an emitter coupled to a cavity.
While commonly used in the literature, there exist several different definitions which obfuscate its utility as a cross-platform metric.
Specifically in the case of non-ideal emitters, the Purcell enhancement can be defined in ways that implicitly depend on the emitter properties if one does not account for additional loss mechanisms.
Here we specify a particular definition of the Purcell enhancement which quantifies the modification of the relevant transition rate and then clarify its relation to other figures of merit used in the literature.

We define the Purcell enhancement $F$ as the ratio of the cavity-enhanced and intrinsic (e.g. Eq. \ref{eq:wigner_weisskopf}) emission rates along the particular transition coupled to the cavity.
In general, $\gamma \neq \gamma_{eg}$, even in an open cavity, as $\gamma$ will include alternative decay channels.
Let $\gamma_{eg,\perp}$ be the contribution to $\gamma$ corresponding to decay through the relevant transition ($\ket{e}\to\ket{g}$) but with emission outside of the cavity.
The Purcell enhancement is then defined as
\begin{equation} \label{eq:purcell_enhance_def}
    F \equiv \frac{\gamma_{eg,\perp}}{\gamma_{eg}}
    + \frac{4g^2}{\kappa \gamma_{eg}} \frac{(\kappa/2)^2}{(\kappa/2)^2 + \delta^2}.
\end{equation}
The advantage of this definition is that it is essentially emitter independent --- the cavity parameters and emitter orientation/placement determine $F$.
We can see this by noting that 
\begin{equation}
    g = \frac{|\mu|}{n}\sqrt{\frac{\omega}{4\pi\epsilon_0 \hbar V}} \left(\frac{\vb E \cdot \mu}{|\vb E_{\text{max}}||\mu|} \right)
\end{equation}
where $\vb E / |\vb E_{\text{max}}|$ is the vector electric field profile of the cavity mode normalized to a maximum of unity and $V$ is the corresponding mode volume.
Using this and Eq. \ref{eq:wigner_weisskopf} we can write the Purcell enhancement as
\begin{equation}
    F = \frac{\gamma_{eg,\perp}}{\gamma_{eg}}
    + F_P
    \left(\frac{\vb E \cdot \mu}{|\vb E_{\text{max}}||\mu|} \right)
    \frac{(\kappa/2)^2}{(\kappa/2)^2 + \delta^2}.
\end{equation}
where
\begin{equation}
    F_P \equiv \frac{3}{4\pi^2} \left(\frac{\lambda}{n}\right)^3 \frac{Q}{V}
\end{equation}
is the well-known \textit{Purcell factor} for a cavity with quality factor $Q=\omega_c / \kappa$.

From this analysis we see that the Purcell enhancement is a metric that describes the effect of the cavity on the emitter, encapsulating both the effectiveness of the coupling (spatially and spectrally) where as the Purcell factor is a figure of merit for the cavity itself, independent of any emitter.
These parameters are related with $F_P$ essentially giving the maximum Purcell enhancement that one could expect from a cavity.
From an engineering standpoint, it is clear that the Purcell enhancement $F$ is the correct metric for comparing designs given that optimization of both coupling and cavity properties is needed.

\subsection{Estimation of Purcell enhancement from experiment}
Estimation of the Purcell enhancement is challenging, especially when additional loss channels are significant.
Using time-resolved spectroscopy, one is able to resolve the temporal profile of emitted photons which follows the excited state population.
This allows one to determine the decay rate $\gamma_{\text{eff}}$ as a function of detuning $\delta$ as in Eq. \ref{eq:gamma_eff}, which is discussed in the main text as the excited state lifetime $\tau \equiv 1 / \gamma_{\text{eff}}$.

In general, $\gamma$ (and likewise $\gamma_{\text{eff}}$) contains contributions from a variety of different mechanisms including the aforementioned emission out of the cavity $\gamma_{eg,\perp}$, radiative transitions along other channels $\gamma_{\text{rad,other}}$ (e.g. via the phonon-sideband), and non-radiative transitions $\gamma_{\text{non-rad}}$.
These contributions are additive with $\gamma = \gamma_{eg,\perp}+\gamma_{\text{rad,other}}+\gamma_{\text{non-rad}}$.
From the analysis, the effect of the cavity is seen to be an additional decay channel with rate $(4g^2/\kappa)$ for $\delta=0$.
When $|\delta| \gg \kappa$ the cavity contribution is suppressed.

We consider the case of an open cavity ($\gamma_{eg,\perp} \approx \gamma_{eg}$) and look at the ratio of the experimentally measured lifetimes for the resonant ($\delta = 0$) and non-resonant ($|\delta| \to \infty$) cases which is given by
\begin{equation}
    \frac{\tau_{\text{off-res}}}{\tau_{\text{on-res}}}
    = \frac{\gamma + 4g^2 / k\kappa}{\gamma} = 1 + \frac{4g^2}{\kappa\gamma}.
\end{equation}
Comparing this to the definition of the Purcell enhancement (Eq. \ref{eq:purcell_enhance_def}) for an open cavity at zero detuning, $F=1+4g^2/\kappa\gamma_{eg}$, we find
\begin{equation}
    F = 1 + \frac{\gamma}{\gamma_{eg}} \left( \frac{\tau_{\text{off-res}}}{\tau_{\text{on-res}}} - 1 \right)
\end{equation}
reproducing the result of \cite{zhang2018strongly} with the correction of an additive constant of unity.

The factor $\gamma/\gamma_{eg}$ is the reciprocal of the fraction of decay via the dipole transition for an intrinsic SiV in bulk diamond.
For an ideal two-level system $\gamma/\gamma_{eg}=1$ and so the Purcell enhancement simply reduces to the ratio of the lifetimes $F = \tau_{\text{off-res}} / \tau_{\text{on-res}}$ which is the expected result.
In general however, $\gamma/\gamma_{eg} > 1$ corresponding to the fact that realistic emitters do not decay entirely through a single transition.
In such cases, estimation of this factor from the literature is necessary and can be accomplished given the branching ratio of the zero-phonon line (ZPL) $\xi\equiv \gamma_{eg} / \gamma_{\text{ZPL}}$, Debye-Waller factor $DW \equiv \gamma_{\text{ZPL}} / (\gamma_{eg} + \gamma_{\text{rad,other}})$, and quantum efficiency $QE\equiv (\gamma_{eg} + \gamma_{\text{rad,other}}) / \gamma$ with $\gamma_{\text{ZPL}}$ being the rate of emission through the ZPL. From these parameters one finds
\begin{equation}
    \frac{\gamma}{\gamma_{eg}} = \frac{1}{\xi \cdot DW \cdot QE}.
\end{equation}
This factor can then be used to estimate the Purcell enhancement as in the main text.

\subsection{Simulation of Purcell enhancement}
A useful feature of our definition of the Purcell enhancement, as in Eq. \ref{eq:purcell_enhance_def}, is that it can be directly simulated using finite-difference-time-domain (FDTD) simulation (Lumerical FDTD, MEEP).
Such methods typically involve approximating the emitter as a dipole antenna driven with a known power and then comparing this power to the actual power radiated in the simulation.
Since both the radiated power and physical emitter dipole transition rate depends on the local density of states, the relative enhancement of the radiated power is itself the Purcell enhancement as in Eq. \ref{eq:purcell_enhance_def}, provided the simulation remains in the leaky- and open-cavity limit.

The Purcell enhancement values as shown in Fig. 1f,g in the main text were computed using this simulation method in Lumerical FDTD.
For these simulations the parameters $(a_{\mathrm{mir}}, a_{\mathrm{cav}}, h_x, h_y, w) = (143, 133, 71, 183, 360)$ were used with the slight deviation of $a_{\mathrm{mir}}$ taken to handle meshing issues at lower resolutions.
The design under consideration utilized 6 mirror holes and 20 taper holes, matching the device considered in the main text.

\begin{figure*}[t]
\centering
\includegraphics[width=\textwidth]{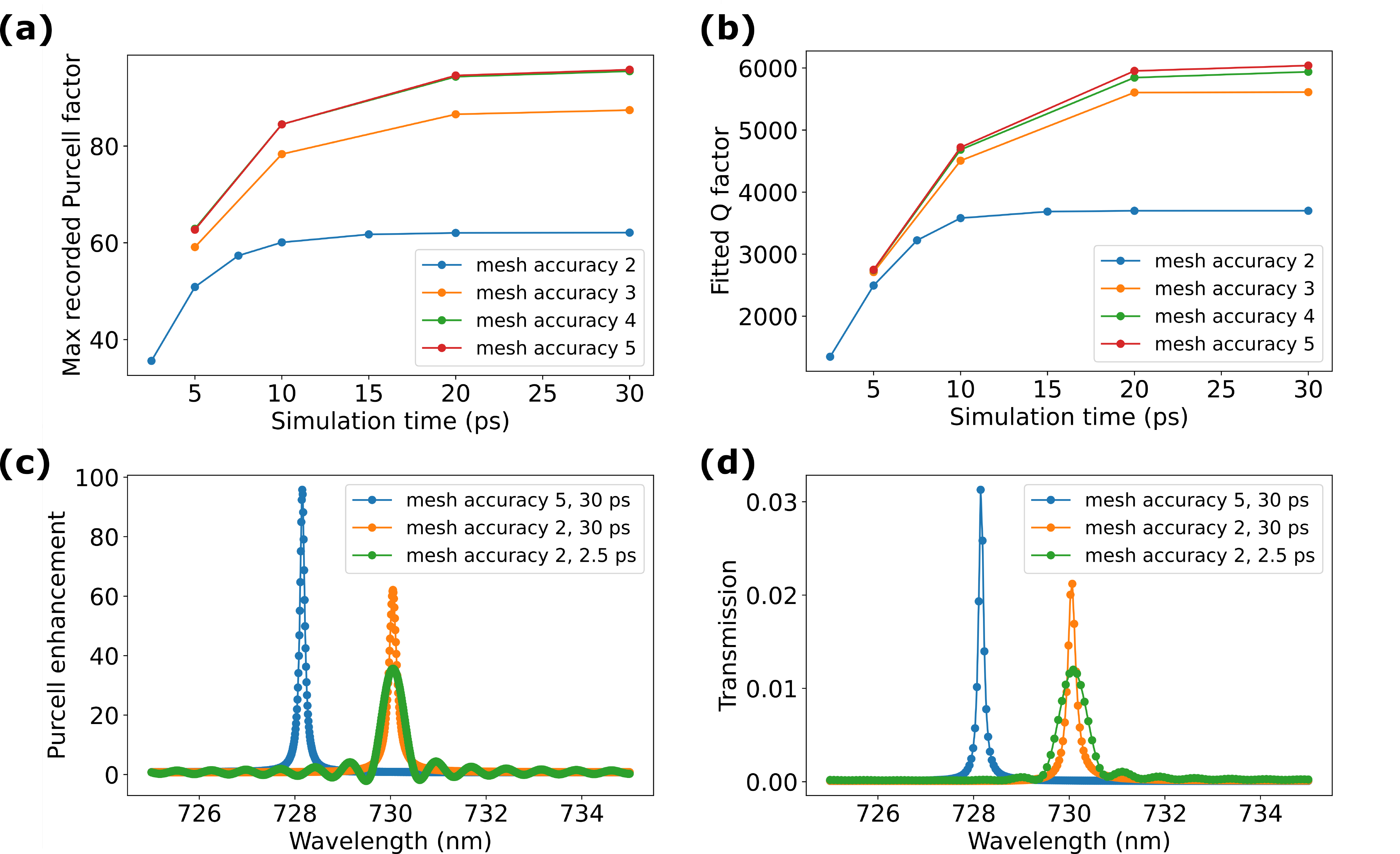}
\caption{\textbf{a--b.} Convergence of simulation parameters with increased mesh accuracy and simulation time. Simulations shown in the main text are done with a mesh accuracy of 4 and simulation time of 2 ps to optimize between accuracy and computation time. \textbf{c--d.} Example spectra corresponding to the indicated points in (\textbf{a--b}). Oscillatory behavior of the spectra at short simulation times is due to the finite bandwidth of the simulation. The resonance frequency shifts with improved mesh resolution. Purcell enhancement and Q are extracted from the spectra.} 
\label{fig:SI_9_1}
\end{figure*}

The convergence of both the Purcell enhancement and cavity quality factor $Q$ as a function of simulation time and mesh size (via the ``mesh accuracy'' setting within Lumerical) is verified as shown in Fig.~\ref{fig:SI_9_1}a,b.
These simulations were performed with an ideally oriented dipole (axis orthogonal to the cavity) positioned 20\,nm below the GaP-diamond interface. 
Incidentally, this also shows the potential of this device given proper orientation of the SiV --- here an increase in the Purcell enhancement by a factor of 4.
The simulations ultimately used in Fig. 1f-g were done with a mesh accuracy of 4 and simulation time of $20$\,ps in order to maintain high accuracy with practical computation time.

Typical Purcell enhancement and transmission spectra obtained from the simulation results are shown in Fig.~\ref{fig:SI_9_1}c,d.
The simulated $Q$ factors were determined by Lorentzian fits to the transmission spectra.
Significant oscillations can be observed in both the Purcell enhancement and transmission spectra for shorter simulation times which is characteristic of a sinc-lineshape resulting from the rectangular windowing of the finite simulation time.
Additionally, the wavelength position of the resonance converges with higher mesh accuracy due to meshing errors.

\newpage
\section{Fits and error approximation}
Fit parameters were determined by a non-linear least-squares fitting algorithm as implemented in Python's SciPy module. 
The fit parameters and corresponding covariance matrix are returned and used for subsequent calculations.
Error bounds are given by the standard deviations which are calculated as the square root of the respective diagonal element of the parameter covariance matrix.

\subsection{Lifetime measurements}
To fit the lifetime data we use an exponentially modified Gaussian (exGaussian) lineshape which is equivalent to a Gaussian impulse response (of the overall system) convolved with a single-sided exponential decay, i.e.
\begin{equation}
    f(t) = \exp(-\frac{(t-\mu)^2}{2\sigma^2}) * \Big(a \exp(-\Gamma t) H(t) \Big),
\end{equation}
where $\mu$ and $\sigma$ correspond to the center position (delay of system response after trigger) and width of the impulse response.
Likewise, $a$, and $\Gamma$ are the amplitude and decay rate of the exponential cavity-defect response.
The function $H(t)$ is the unit-step (Heaviside-step) function which is zero for $t<0$ and unity for $t>0$.
An explicit form for the exGaussian lineshape can be computed directly
\begin{equation}
    f(t) = a \int_0^\infty \dd{\tau} \exp(-\frac{(t - \tau-\mu)^2}{2\sigma^2}) \exp(-\Gamma \tau).
\end{equation}
Changing variables to $z = (\tau - (t - \mu - \sigma^2 \Gamma)) / \sqrt{2\sigma^2}$ gives
\begin{equation}
    f(t) = a \exp(-\Gamma\left( (t-\mu) - \frac{\sigma^2\Gamma}{2} \right)) \int_x^\infty (\sqrt{2\sigma^2}\dd{z}) e^{-z^2}
\end{equation}
where $x = -((t-\mu) - \sigma^2 \Gamma)/\sqrt{2\sigma^2}$.
The second term is identified as the complementary error function $\mathrm{erfc}(x) \equiv (2/\sqrt{\pi})\int_x^\infty \dd{z} e^{-z^2}$.
Thus, the final lineshape is given by
\begin{equation}
    f(t) = a\sigma \sqrt{\frac{\pi}{2}} \exp(-\Gamma\left( (t-\mu) - \frac{\sigma^2\Gamma}{2} \right) )
    \mathrm{erfc}\left( -\frac{(t-\mu) - \sigma^2 \Gamma}{\sqrt{2\sigma^2}} \right),
\end{equation}
which was used for fits of the lifetime data.
In fitting the actual data some additional modifications needed to be made.
Firstly, the width of the Gaussian input response ($\sigma$) is determined to be 228\,ps from a fit of the instrument impulse response which was measured using a reflection of the 2-ps pulse of the driving laser (which is thus far below the resolution of the timing board $\sim80$-ps).
Secondly, an additional free parameter $b$ is needed to account for the background (dark counts, light leakage, etc.) of the system such that the overall fitting function is given by $f^\prime(t) \equiv f(t) + b$.

As shown above, the original data fits the decay constant $\Gamma$ of the population decay which is given in units of inverse time.
The population lifetime is defined as $\tau \equiv 1/\Gamma$.
With the (standard-deviation) error of $\Gamma$, $\sigma_\Gamma$ computed from the self-variance of the fit as described above, the error in the lifetime is given from a Taylor expansion as
\begin{equation}
    \sigma_\tau = \abs{\frac{\sigma_\Gamma}{\Gamma^2}}.
\end{equation}
The lifetimes and corresponding errors are tabulated below in table \ref{tab:lifetime_fits_stddev}.

\begin{table}[ht]

    \caption{\label{tab:lifetime_fits_stddev}%
        Standard deviations of fitted lifetimes.
    }
    \begin{ruledtabular}
        \begin{tabular}{cccccc}
        \textrm{SiV}&
        \textrm{Meas. No.}&
        $\Gamma$ (ns$^{-1}$)&
        $\sigma_\Gamma$ (ns$^{-1}$)&
        $\tau$ (ns)&
        $\sigma_\tau$ (ns)
        \\
        
    \colrule
        
        \textrm{P-1}
         & 1  & 0.644 & 0.025 & 1.55 & 0.06 \\
         & 2  & 0.577 & 0.020 & 1.73 & 0.06 \\
         & 3  & 0.615 & 0.012 & 1.63 & 0.03 \\
         & 4  & 0.656 & 0.010 & 1.52 & 0.02 \\
         & 5  & 0.799 & 0.007 & 1.25 & 0.01 \\
         & 6  & 0.887 & 0.009 & 1.13 & 0.01 \\
         & 7  & 0.716 & 0.009 & 1.40 & 0.02 \\
         & 8  & 0.718 & 0.006 & 1.39 & 0.01 \\
         & 9  & 0.661 & 0.011 & 1.51 & 0.02 \\
         & 10 & 0.591 & 0.008 & 1.69 & 0.02 \\
         & 11 & 0.573 & 0.015 & 1.75 & 0.05 \\
         & 12 & 0.569 & 0.011 & 1.76 & 0.03 \\
         & 13 & 0.527 & 0.014 & 1.90 & 0.05 \\
         & 14 & 0.517 & 0.015 & 1.94 & 0.06 \\

    \colrule
        \textrm{P-2}
         & 1 & 0.675 & 0.023 & 1.48 & 0.05 \\
         & 2 & 0.846 & 0.013 & 1.18 & 0.02 \\
         & 3 & 1.034 & 0.013 & 0.97 & 0.01 \\
         & 4 & 1.384 & 0.015 & 0.72 & 0.01 \\
         & 5 & 2.010 & 0.022 & 0.50 & 0.01 \\
         & 6 & 2.030 & 0.034 & 0.49 & 0.01 \\
         & 7 & 1.703 & 0.022 & 0.59 & 0.01 \\
         & 8 & 1.753 & 0.032 & 0.57 & 0.01 \\
         & 9 & 1.816 & 0.040 & 0.55 & 0.01 \\
        
        \end{tabular}
    \end{ruledtabular}

\end{table}

To compute the propagated error of the lifetime ratio (such as $\tau_{\text{off-res}}/\tau_{\text{on-res}}$ as in equation (2) of the main text), we assume statistical independence of the lifetime decay constants (i.e. the covariance is $0$).
The error propagation of the lifetime ratio
\begin{equation}
    r = \frac{\tau_1}{\tau_2}
\end{equation}
can be computed from a Taylor expansion as
\begin{equation}
    \sigma_r = \abs{r}\sqrt{
    \left( \frac{\sigma_{\tau_1}}{\tau_1}\right)^2 
    +
    \left( \frac{\sigma_{\tau_2}}{\tau_2}\right)^2
    }.
\end{equation}
The table \ref{tab:ratio_fit_stddev} shows the standard deviation of the lifetime parameter fits for SiV P-1 and P-2, as well as the corresponding error of the lifetime ratio.

\begin{table}[ht]

    \caption{\label{tab:ratio_fit_stddev}%
        Standard deviations of lifetime ratio parameters.
    }
    \begin{ruledtabular}
        \begin{tabular}{ccccccc}
        \textrm{SiV}&
        $\tau_{\text{off-res}}$ (ns)&
        $\sigma_{\tau,\text{off-res}}$ (ns)&
        $\tau_{\text{on-res}}$ (ns)&
        $\sigma_{\tau,\text{on-res}}$ (ns)&
        $r=\tau_{\text{off-res}}/\tau_{\text{on-res}}$&
        $\sigma_r$
        \\
        \colrule
        \textrm{P-1} &
        1.94 &
        0.06 &
        1.13 &
        0.01 &
        1.72 &
        0.05 \\
        \textrm{P-2} &
        1.48 &
        0.05 &
        0.50 &
        0.01 &
        2.98 &
        0.11 \\
        \end{tabular}
    \end{ruledtabular}
    
\end{table}

\subsection{Resonant excitation measurements: linewidth and cooperativity}
The linewidth of the SiV transitions are extracted from the resonant excitation measurements by fitting to a Lorentzian lineshape given by
\begin{equation}
    g(t) = \frac{(\Delta\nu/2)^2}{(\nu-\nu_0)^2 + (\Delta\nu/2)^2},
\end{equation}
where $\Delta\nu$ gives the full-width-half-max (FWHM) in dimensions of $\nu$ (here in units of Hertz; note that decay constants $\gamma$ from the Lindblad operators are given by $\gamma \equiv 2\pi\Delta\nu$).
As before, the error of the linewidth $\sigma_{\Delta\nu}$ is given by the square-root of the corresponding diagonal element from the fit covariance matrix.
The corresponding error bounds are directly referenced in Figure 5 of the main text.

In general, the measured linewidth from the resonant excitation measurements are broadened by spectral diffusion and pure dephasing.
We would like to compare this to the lifetime-limited linewidth which is determined by the Fourier transform of the excited state \textit{amplitude} decay.
Assuming a form of the excited-state amplitude decay $x(t) = \exp(-\Gamma t /2) H(t)$ (which corresponds to the population decay $|x(t)|^2$ with rate $\Gamma$), the Fourier transform in angular-frequency space is
\begin{equation}
    x(t)= \exp(-\Gamma t /2) H(t) \quad\xrightarrow{\mathrm{F.T.}}\quad \tilde x (\omega) = \frac{1}{(\Gamma/2) + i\omega}
\end{equation}
where $\omega \equiv 2\pi \nu$ is the angular frequency for an oscillation at frequency $\nu$ in Hertz.
The lineshape of a lifetime-limited transition follows the spectral density $|\tilde x (\omega)|^2$ which has Lorentzian profile with FWHM in $\omega$-space of $\Gamma$.
Correspondingly, the FWHM in $\nu$-space is then $\Gamma/2\pi$, or correspondingly $1/2\pi \tau$ in terms of the lifetime.
The associated error is then $\sigma_{\text{FWHM},\tau} = |\sigma_{\tau}/2\pi\tau^2|$.
Error bounds for lifetime-limited linewidth in the on-/off-resonant cases are shown below in table \ref{tab:linewidth_lifetime_limit_stddev}.

The cooperativity is given by $C = 4g^2/\kappa(\gamma + \gamma_d)$ where $\gamma_d$ describes the contribution to the loss via pure dephasing of the SiV linewidth (see Section \ref{sec:model} for additional details).
The measured linewidth in the resonant excitation measurements gives the broadened linewidth which we associate to the total loss $\gamma_{\text{tot}} \equiv \gamma + \gamma_d$ (in $\omega$-space) whose errors are tablulated in table \ref{tab:linewidth_lifetime_limit_stddev}.
As discussed in the main text, in the limit of negligible pure dephasing ($\gamma_d \to 0$), we have
\begin{equation}
    C_{\text{max}} = \frac{\tau_{\text{off-res}}}{\tau_{\text{on-res}}} - 1.
\end{equation}
As a result the error bounds on $C_{\text{max}}$ correspond to that of the lifetime ratios themselves.
In order to estimate the actual cooperativity given the measured linewidth we use the following relation
\begin{equation}
    C = \frac{\gamma}{\gamma_{\text{tot}}} C_{\text{max}} 
\end{equation}
where $\gamma = 1/2\pi\tau_{\text{off-res}}$ is numerically given by the off-resonant lifetime-limited linewidth as tabulated in table \ref{tab:linewidth_lifetime_limit_stddev}.
We first compute the error of the linewidth ratio assuming statistical independence of each of the parameters such that the overall error can be computed
\begin{equation}
    \sigma_{\gamma/\gamma_{\text{tot}}} = \frac{\gamma}{\gamma_{\text{tot}}} 
    \sqrt{\left( \frac{\sigma_{\gamma}}{\gamma} \right)^2 + 
    \left( \frac{\sigma_{\gamma_{\text{tot}}}}{\gamma_{\text{tot}}} \right)^2}.
\end{equation}
The error associated with the cooperativity is thus
\begin{equation}
    \sigma_C = C \sqrt{
    \left( \frac{\sigma_{\gamma/\gamma_{\text{tot}}}}{\gamma/\gamma_{\text{tot}}} \right)^2 + 
    \left( \frac{\sigma_{C_{\text{max}}}}{C_{\text{max}}} \right)^2}
\end{equation}
The calculated values are tabulated in table \ref{tab:linewidth_lifetime_limit_stddev}.

\begin{table}[ht]

    \caption{\label{tab:linewidth_lifetime_limit_stddev}%
        Standard deviations of parameters for cooperativity estimation. All frequencies (rates) are in units of MHz.
    }

        \begin{ruledtabular}
        \begin{tabular}{ccccccccccc}
        \textrm{SiV} & 
        $\gamma/2\pi$ &
        $\sigma_{\gamma}/2\pi$ & 
        $\gamma_{\text{tot}}/2\pi$ & 
        $\sigma_{\gamma_\text{tot}}/2\pi$ &
        $\gamma/\gamma_{\text{tot}}$ &
        $\sigma_{\gamma/\gamma_{\text{tot}}}$ &
        $C_{\text{max}}$ &
        $\sigma_{C_{\text{max}}}$ &
        $C$ &
        $\sigma_{C}$ \\
        \colrule
        P-1 &
            82.22 &
            2.39 &
            578.81 &
            31.90 &
            0.140 &
            0.009 &
            0.72 &
            0.05 &
            0.10 &
            0.01 
        \\
        P-2 &
            107.42 &
            3.64 &
            - &
            - &
            - &
            - &
            1.98 &
            0.10 &
            - &
            - 
        \end{tabular}
    \end{ruledtabular}

\end{table}

\bibliography{supplement}